\documentclass[superscriptaddress,showpacs,nofootinbib,aps,prd,twocolumn]{revtex4}
\usepackage{epsfig}
\usepackage{graphicx}
\usepackage{dcolumn}
\usepackage{bm}

\date{}
\begin{document}


\newcommand{\ds}{\displaystyle}
\newcommand{\mc}{\multicolumn}
\newcommand{\bce}{\begin{center}}
\newcommand{\ece}{\end{center}}
\newcommand{\beq}{\begin{equation}}
\newcommand{\eeq}{\end{equation}}
\newcommand{\bea}{\begin{eqnarray}}

\newcommand{\eea}{\end{eqnarray}}
\newcommand{\cont}{\nonumber\eea\bea}
\newcommand{\cl}[1]{\begin{center} {#1} \end{center}}
\newcommand{\ba}{\begin{array}}
\newcommand{\ea}{\end{array}}

\newcommand{\ab}{{\alpha\beta}}
\newcommand{\cd}{{\gamma\delta}}
\newcommand{\dc}{{\delta\gamma}}
\newcommand{\ac}{{\alpha\gamma}}
\newcommand{\bd}{{\beta\delta}}
\newcommand{\abc}{{\alpha\beta\gamma}}
\newcommand{\eps}{{\epsilon}}
\newcommand{\lam}{{\lambda}}
\newcommand{\mn}{{\mu\nu}}
\newcommand{\mpnp}{{\mu'\nu'}}
\newcommand{\Amuu}{{A_{\mu}}}
\newcommand{\Amuo}{{A^{\mu}}}
\newcommand{\Vmuu}{{V_{\mu}}}
\newcommand{\Vmuo}{{V^{\mu}}}
\newcommand{\Anuu}{{A_{\nu}}}
\newcommand{\Anuo}{{A^{\nu}}}
\newcommand{\Vnuu}{{V_{\nu}}}
\newcommand{\Vnuo}{{V^{\nu}}}
\newcommand{\Fmnu}{{F_{\mu\nu}}}
\newcommand{\Fmno}{{F^{\mu\nu}}}

\newcommand{\abcd}{{\alpha\beta\gamma\delta}}


\newcommand{\bsigma}{\mbox{\boldmath $\sigma$}}
\newcommand{\btau}{\mbox{\boldmath $\tau$}}
\newcommand{\brho}{\mbox{\boldmath $\rho$}}
\newcommand{\bpipi}{\mbox{\boldmath $\pi\pi$}}
\newcommand{\bss}{\bsigma\!\cdot\!\bsigma}
\newcommand{\btt}{\btau\!\cdot\!\btau}
\newcommand{\bnabla}{\mbox{\boldmath $\nabla$}}
\newcommand{\bphi}{\mbox{\boldmath $\tau$}}
\newcommand{\bvarphi}{\mbox{\boldmath $\rho$}}
\newcommand{\bDelta}{\mbox{\boldmath $\Delta$}}
\newcommand{\bpsi}{\mbox{\boldmath $\psi$}}
\newcommand{\bPsi}{\mbox{\boldmath $\Psi$}}
\newcommand{\bPhi}{\mbox{\boldmath $\Phi$}}
\newcommand{\bnab}{\mbox{\boldmath $\nabla$}}
\newcommand{\bpi}{\mbox{\boldmath $\pi$}}
\newcommand{\btheta}{\mbox{\boldmath $\theta$}}
\newcommand{\bkappa}{\mbox{\boldmath $\kappa$}}

\newcommand{\bA}{{\bf A}}
\newcommand{\bB}{\mbox{\boldmath $B$}}
\newcommand{\bp}{\mbox{\boldmath $p$}}
\newcommand{\bk}{\mbox{\boldmath $k$}}
\newcommand{\bc}{\mbox{\boldmath $c$}}
\newcommand{\bq}{\mbox{\boldmath $q$}}
\newcommand{\bfe}{{\bf e}}
\newcommand{\bb}{\mbox{\boldmath $b$}}
\newcommand{\br}{\mbox{\boldmath $r$}}
\newcommand{\bR}{\mbox{\boldmath $R$}}

\newcommand{\fph}{${\cal F}$}
\newcommand{\aph}{${\cal A}$}
\newcommand{\dph}{${\cal D}$}
\newcommand{\fpi}{f_\pi}
\newcommand{\mpi}{m_\pi}
\newcommand{\Tr}{{\mbox{\rm Tr}}}
\def\Qb{\overline{Q}}
\newcommand{\delu}{\partial_{\mu}}
\newcommand{\delo}{\partial^{\mu}}
%
%
\newcommand{\up}{\!\uparrow}
\newcommand{\upup}{\uparrow\uparrow}
\newcommand{\updo}{\uparrow\downarrow}
\newcommand{\uu}{$\uparrow\uparrow$}
\newcommand{\ud}{$\uparrow\downarrow$}
\newcommand{\auu}{$a^{\uparrow\uparrow}$}
\newcommand{\aud}{$a^{\uparrow\downarrow}$}
\newcommand{\pu}{p\!\uparrow}

\newcommand{\qp}{quasiparticle}
\newcommand{\sa}{scattering amplitude}
\newcommand{\ph}{particle-hole}
\newcommand{\qcd}{{\it QCD}}
\newcommand{\integ}{\int\!d}
\newcommand{\ie}{{\sl i.e.~}}
\newcommand{\etal}{{\sl et al.~}}
\newcommand{\etc}{{\sl etc.~}}
\newcommand{\rhs}{{\sl rhs~}}
\newcommand{\lhs}{{\sl lhs~}}
\newcommand{\eg}{{\sl e.g.~}}
\newcommand{\ef}{\epsilon_F}
\newcommand{\sigt}{d^2\sigma/d\Omega dE}
\newcommand{\sige}{{d^2\sigma\over d\Omega dE}}
\newcommand{\rpaeq}{\beq
\left ( \begin{array}{cc}
A&B\\
-B^*&-A^*\end{array}\right )
\left ( \begin{array}{c}
X^{(\kappa})\\Y^{(\kappa)}\end{array}\right )=E_\kappa
\left ( \begin{array}{c}
X^{(\kappa})\\Y^{(\kappa)}\end{array}\right )
\eeq}
\newcommand{\ket}[1]{| {#1} \rangle}
\newcommand{\bra}[1]{\langle {#1} |}
\newcommand{\ave}[1]{\langle {#1} \rangle}
\newcommand{\half}{{1\over 2}}

\newcommand{\singlespace}{
    \renewcommand{\baselinestretch}{1}\large\normalsize}
\newcommand{\doublespace}{
    \renewcommand{\baselinestretch}{1.6}\large\normalsize}
\newcommand{\bftau}{\mbox{\boldmath $\tau$}}
\newcommand{\bfalpha}{\mbox{\boldmath $\alpha$}}
\newcommand{\bfgamma}{\mbox{\boldmath $\gamma$}}
\newcommand{\bfxi}{\mbox{\boldmath $\xi$}}
\newcommand{\bfbeta}{\mbox{\boldmath $\beta$}}
\newcommand{\bfeta}{\mbox{\boldmath $\eta$}}
\newcommand{\bfpi}{\mbox{\boldmath $\pi$}}
\newcommand{\bfphi}{\mbox{\boldmath $\phi$}}
\newcommand{\bfR}{\mbox{\boldmath ${\cal R}$}}
\newcommand{\bfL}{\mbox{\boldmath ${\cal L}$}}
\newcommand{\bfM}{\mbox{\boldmath ${\cal M}$}}
\def\dblint{\mathop{\rlap{\hbox{$\displaystyle\!\int\!\!\!\!\!\int$}}
    \hbox{$\bigcirc$}}}
\def\ut#1{$\underline{\smash{\vphantom{y}\hbox{#1}}}$}

\def\UNITY{{\bf 1\! |}}
\def\Pom{{\bf I\!P}}
\def\lsim{\mathrel{\rlap{\lower4pt\hbox{\hskip1pt$\sim$}}
    \raise1pt\hbox{$<$}}}         
\def\gsim{\mathrel{\rlap{\lower4pt\hbox{\hskip1pt$\sim$}}
    \raise1pt\hbox{$>$}}}         
\def\beq{\begin{equation}}
\def\eeq{\end{equation}}
\def\bea{\begin{eqnarray}}
\def\eea{\end{eqnarray}}

{\large FZJ-IKP-TH-2006-11}\\

\title{Quenching of Leading Jets and Particles: 
the $p_\perp$ Dependent Landau-Pomeranchuk-Migdal effect
from  Nonlinear $k_\perp$ Factorization}

\author{N.N. Nikolaev}%
\email{N.N. Nikolaev@fz-juelich.de}
\affiliation{Institut f\"ur Kernphysik, Forschungszentrum J\"ulich, D-52425 J\"ulich, Germany}
\affiliation{L.D. Landau Institute for Theoretical Physics, 142432 Chernogolovka, Russia}
\author{W. Sch\"afer}%
\email{Wo.Schaefer@fz-juelich.de}
\affiliation{Institute of Nuclear Physics PAN, PL-31-342 Cracow, Poland}

\begin{abstract}
We report the first derivation of radiative nuclear stopping
(the  Landau-Pomeranchuk-Migdal
effect) for leading jets at fixed values of the transverse momentum $\bp$
in the beam fragmentation region of hadron-nucleus collisions from
RHIC (Relativistic Heavy Ion Collider) to LHC (Large Hadron
Collider). The major novelty
of this work is a derivation of the missing virtual radiative pQCD correction
to these processes  - the real-emission radiative
corrections are already available
in the literature. We manifestly implement the unitarity
relation, which in the simplest form requires that upon summing over
the virtual and real-emission corrections the total number of scattered
quarks must exactly equal unity. For the free-nucleon target,
the leading jet spectrum is shown to satisfy
the familiar linear Balitsky-Fadin-Kuraev-Lipatov leading log-${1\over x}$
(LL${1\over x}$) evolution.
For nuclear targets,  the nonlinear $k_\perp$-factorization
for the LL-${1\over x}$ evolution of the leading jet spectrum
is shown to exactly match the equally nonlinear
LL${1\over x}$ evolution of the collective nuclear glue -- there emerges
a unique linear $k_\perp$-factorization relation between the two
nonlinear evolving nuclear observables.
We argue that within the standard dilute
uncorrelated nucleonic gas treatment of heavy nuclei, in the finite
energy range from RHIC to LHC, the
leading jet spectrum can be evolved
in the LL${1\over x}$ Balitsky-Kovchegov approximation.
We comment on the extension of these
results to, 
and their possible reggeon field theory interpretation for,
 mid-rapidity jets. 
\end{abstract}
\date{\today}
\maketitle



\section{Introduction}
In this communication we address two related issues in the
 perturbative Quantum
Chromo Dynamics (pQCD)
description of the single-jet inclusive
spectra in hadron-nucleon and hadron-nucleus 
collisions. Our principal interest is in the 
nonlinear $k_\perp$ factorization for 
the pQCD radiative quenching (stopping) of leading jets with fixed
transverse momentum $\bp$ produced at 
large values of the Feynman variable $x_F$, i.e., in the beam 
fragmentation region of hadron-nucleon and hadron-nucleus collisions.
As it has become customary for the nuclear dependence of radiative
effects, we refer to nuclear stopping of leading jets at
fixed $\bp$ as the 
$\bp$-dependent Landau-Pomeranchuk-Migdal 
(LPM) effect. 
There is an extensive literature on the LPM effect for mid-rapidity 
jets, where the prime concern is in the radiation induced 
loss of the transverse, with respect to the hadron-nucleus 
collision axis, momentum 
of the jet when one integrates over the transverse
momentum of the radiation with respect to the jet axis
(\cite{SlavaLPM}, for reviews and further references see 
\cite{SlavaReview,KovchegovReview}). In the mid-rapidity kinematics, 
the transverse momentum of radiation with respect to the parent 
high-$p_\perp$ parton only amounts to a small shift of the (pseudo)rapidity 
of the high-$p_\perp$ parton which is hardly observable due to
the boost invariance in the mid-rapidity region. In contrast to that,
leading jets with $x_F \sim 1$
are generated by hard scattering of valence quarks.
The density of valence quarks drops rapidly 
at large values of the Bjorken variable in the beam nucleon,
$x_N \to 1$, which entails that  the 
$x_F$ dependence of the leading jet cross 
section is extremely sensitive 
to the radiative energy loss. We develop a 
nonlinear $k_\perp$ factorization description of leading jet
production with full allowance for the real and virtual QCD
radiative corrections. In the second part we focus on the
leading log${1\over x}$ (LL${1\over x}$)
 properties of the leading jet
spectrum. We argue that this spectrum offers a 
long sought linear mapping of the unintegrated collective nuclear 
glue and comment on the extension of this finding to 
mid-rapidity jets.
 
Within pQCD,  there are several subtleties in the application of 
factorization theorems
to the leading quark production in $qN, qA$ collisions. 
In the more familiar collinear approximation
described in all textbooks \cite{Textbook}, 
the relevant pQCD subprocess is hard scattering of valence quarks
off soft gluons with $x \ll 1$ in the target, which can also be
viewed as a hard breakup (excitation) $q\to q'g$.
All partons enter the hard pQCD scattering with vanishing transverse 
momentum and the large $p_\perp$ of the observed jet is generated 
in the real-emission hard breakup. The inadequacy of such a collinear 
approximation is well known, a recent concise 
discussion of the 
necessity of fully unintegrated parton densities with
explicit allowance for transverse momenta is
found in \cite{CollinsZu}. In the high-energy limit, 
the latter is furnished by the so-called
$k_\perp$-factorization which goes back to 
the classic 1975-1978 
works on the Balitsky-Fadin-Kuraev-Lipatov (BFKL) equation for the
small-$x$ evolution of the unintegrated glue (\cite{KLF,BFKL}, see
also the
recent review on $k_\perp$-factorization and more references in
\cite{Smallx,Antoni}). 
In contrast to the collinear 
factorization, within the $k_\perp$-factorization, the hard 
breakup $q\to q'g$ is but the real-emission pQCD radiative 
correction (REC) to the lower order (Born) subprocess, 
which is the quasielastic 
scattering $qN\to qN^*$, $qA\to qA^*$, where the target debris 
$N^*,A^*$ is left in a color excited state.
Evidently, the real-emission
must be complemented by the matching 
virtual pQCD radiative corrections 
(VRC)
to the lower order quasielastic scattering - the real-emission
and virtual radiative corrections are two integral parts of the
small-$x$ evolution of quasielastic scattering, see
Fig. \ref{fig:VirtualReal}.

\begin{widetext}
\mbox{} 
\begin{figure}[!h]
\begin{center}
\includegraphics[width = 6cm,angle=270]{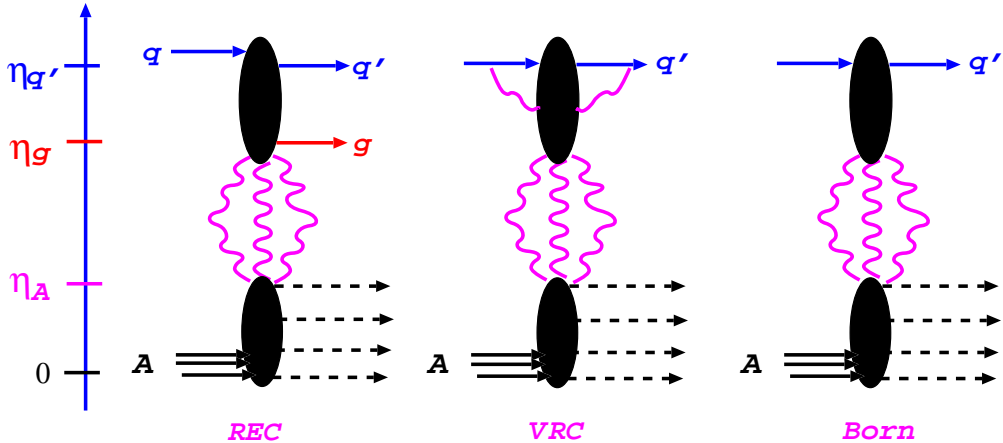}
\caption{\label{fig:VirtualReal}
The real emission (REC) and virtual radiation (VRC) contributions
to the Born spectrum of leading quark jets in inclusive production off
nuclei. The (pseudo)rapidity of radiation correction gluons runs
between the quark jet rapidity and the nuclear boundary condition
rapidity $\eta_A=\log{1\over x_A}$, where $x_A$ is defined by Eq.
(\ref{eq:2.C.1}).}
\end{center}
\end{figure}
\mbox{}
\end{widetext}


Hard scattering off free nucleons is described
by single-gluon exchange in the $t$-cannel. In collisions with nuclei,
the enhancement of multigluon exchanges
leads to a novel concept of nonlinear $k_\perp$ factorization
\cite{Nonlinear,PionDijet,QuarkGluonDijet,Nonuniversality,GluonGluonDijet,Paradigm}.
The Born and VRC amplitudes describe final states in which
the large transverse momentum of the leading jet is compensated by
hadrons at rapidities $\eta \lsim \eta_A$, while in REC such
a compensation is provided also by gluon jets with rapidities
$\eta_A \lsim \eta_g \lsim \eta_{q'}$.
 A theoretical treatment
of the $\bp$ dependent LPM effect for leading jets is necessary
for the interpretation of the forthcoming experimental data on
leading particles from the RHIC to LHC
(\cite{BRAHMS} and references therein).

The interplay of the virtual and real radiative corrections
to single jet spectra is controlled by the
unitarity condition. In its simplest
form unitarity demands that the multiplicity of   
final state leading quarks from the real radiation, and from the
VRC-corrected lower order (Born) quasielastic scattering add up exactly 
to unity. In other words, the small-$x$ evolutions of
the total inclusive cross section for the leading quark 
production and of the total cross section of the quark-target 
interaction must preserve their exact equality.
Evidently, the accurate 
evaluation of the energy loss requires the  
REC for a finite fraction, $z_g$, of the incident
quark's momentum carried by the radiated gluon. Correspondingly,
the VRC must be calculated to a matching accuracy.
While the required real-emission spectra do exist  in
the literature for both the free-nucleon and nuclear targets
(we follow the approach of Ref. \cite{SingleJet}, for 
related works and further references see \cite{Kovchegov,Blaizot,Baier}),
the matching derivation of virtual radiative corrections was
not yet available - it is a major novelty of this paper. Our
derivation of the combined VRC and REC fully respects the
$s$-channel unitarity constraints and paves the way to a first
consistent description of the LPM effect for leading jets 
for fixed $\bp$ of the jet.

Much insight into the interplay of the virtual and real-emission
radiative corrections for nuclear targets comes from the study 
of  LL${1\over x}$ evolution of the LPM effect-deconvoluted 
inclusive spectra
of leading particles, i.e., neglecting the
radiative energy loss. At moderately high energies,
the unitarity constraints in hard 
$qN$ collisions are still
unimportant and they can be described by single-gluon 
exchange in the $t$-channel between 
the projectile and target. In this case a comparison of the 
LL${1\over x}$ evolution of the LPM-effect deconvoluted
leading parton spectrum and of
the BFKL evolution of the unintegrated glue of the nucleon 
leads to a simple $k_\perp$-factorization relationship 
\beq
{d{\sigma}_{Qel} \over d^2\bp} ={1\over 2}f(x,\bp)
\label{eq:1.1}
 \eeq
(for the definition of the unintegrated glue $f(x,\bp)$ in the free
nucleon see below). I.e., the leading quark spectrum
satisfies the same  BFKL small-$x$ evolution equation
as the unintegrated
glue \cite{KLF,BFKL}, and this 
spectrum emerges as a direct probe of the unintegrated glue in the
target nucleon. It is remarkable that our derivation of this 
property makes a direct use the $s$-channel unitarity for
color dipole scattering. In the realm of the BFKL approach, this 
property can be regarded as a straightforward consequence of the
correspondence between LL${1\over x}$ approximation and the dominance 
of multiregge production processes \cite{KLF,BFKL,FadinNew}. 

In contrast to 
the single $t$-channel gluon exchange with the free-nucleon target, 
multiple gluon exchanges with the target nucleus are enhanced by a large 
thickness  of the nucleus. Here the very possibility of describing
a nucleus by a unique collective unintegrated glue, and the
existence of factorization theorems, become problematic. Various
aspects of this issue have been addressed to within the so-called
Color Glass Condensate approach, in which the starting idea was
to describe the nucleus by the collective Weizs\"acker-Williams gluon
density (\cite{CGC} and references therein). More recently, 
much progress has been made within the nonlinear $k_\perp$-factorization
reformulation of the multiple scattering theory for color dipoles. 
Within the color dipole approach, the extensive previous studies of the dijet spectra
(\cite{Nonlinear,PionDijet,QuarkGluonDijet,Nonuniversality,GluonGluonDijet,Paradigm},
see also the recent Ref. \cite{Fujii}) and single-jet spectra \cite{SingleJet}
revealed a striking breaking of the conventional linear $k_\perp$
factorization for hard processes in a nuclear environment. 
Specifically, all the dijet observables can be represented as 
nonlinear quadratures of the collective nuclear glue defined 
in terms of a reference nuclear process - production of 
coherent diffractive dijets \cite{NSSdijet,NSSdijetJETPLett}.
On the other hand, in Refs. \cite{Nonlinear,SingleJet} we noticed the
phenomenon of Abelianization of the single particle spectra --
to LL${1\over x}$ the nonlinear $k_\perp$-factorization simplifies
down to the linear $k_\perp$-factorization subject to the judicious 
choice of the component of the color space density matrix for the
collective nuclear glue. Here we demonstrate that, at least to the
first order of  LL${1\over x}$, a similar Abelianization takes place
for the leading jet spectrum from inelastic processes. 

When speaking of  $k_\perp$-factorization for nuclear collisions,
one needs to define the collective nuclear glue in terms of the
properly chosen physics observable. 
Here we start with the BFKL definition of the
unintegrated glue for free nucleons \cite{KLF,BFKL}.  
As noticed in \cite{NZsplit}, this unintegrated glue can
be directly measured in diffractive production of  
dijets (for other applications of the 
unintegrated glue see the reviews in
\cite{Smallx,Antoni,CollinsZu,CollinsAPP}). 
Generalizing this observation
to nuclear targets \cite{NSSdijet,Nonlinear}, we define 
the collective nuclear glue
in terms of the color-dipole nuclear 
${\textsf S}$-matrix which is accessible experimentally via diffractive 
dijets. 
The LL${1\over x}$ evolution of the so-defined glue is 
a nonlinear one. The REC to the inclusive single leading 
parton spectrum derived in  \cite{SingleJet}
is a nonlinear functional of the collective nuclear 
glue. Equally highly nonlinear is the VRC
correction to quasielastic parton--nucleus scattering derived in this 
paper.  Our nontrivial finding is that 
the leading parton spectrum and the collective glue
share identical first iteration of the nonlinear LL${1\over x}$
evolution, which is the second novelty of our paper. 
One is tempted to  conjecture that
the spectrum of leading jets is linear $k_\perp$-factorizable in
terms of the  collective nuclear glue to all order of the nonlinear
LL${1\over x}$ evolution, i.e., it is a long sought unique linear
probe of the collective nuclear glue. 
It is by now
understood that the closed
form of all-order LL${1\over x}$ evolution for nuclei
does not exist (a detailed review is found in 
\cite{Trianta}).
Still we argue that in view of 
a high accuracy of the dilute
uncorrelated nucleonic gas picture of heavy nuclei, the
small-$x$ evolution of the leading jet spectrum from RHIC to LHC
can to a controlled accuracy be
described by several iterations of the 
Balitsky-Kovchegov (BK) approximation 
\cite{BalitskyNonlinear,KovchegovNonlinear}.

We find it very instructive to compare nonlinear 
$k_\perp$-factorization properties of the  LL${1\over x}$ evolution
of the inclusive spectrum of 
leading quarks belonging to the color-triplet (fundamental)
and color-octet (adjoint) representations of $SU(N_c)$. 
The latter case is of interest from the viewpoint of 
supersymmetric QCD, one can also think of leading gluons
in interactions of glueballs.  A pattern of the nonlinear 
$k_\perp$-factorization for the spectrum changes strikingly
from the triplet to octet
quarks, equally different 
is the nonlinear BFKL evolution of the
components of the color density matrix for the collective glue 
for the triplet-antitriplet
and octet-octet color dipoles. For instance, for strongly absorbing 
nucleus and triplet-antitriplet dipoles, the nonlinear BFKL evolution 
looks as a fusion of two nuclear pomerons, i.e., it is
dominated by triple-pomeron interactions. In striking contrast to
that, for the octet-octet dipoles the dominant term looks
as a fusion of three pomerons from one component of
the color density matrix for the collective nuclear glue
to the fourth pomeron described by a different collective glue.
In both cases, however, 
the evolution of the two quantities -- the leading parton spectrum 
and the collective glue -- turns out to be identical, which lends
more support to our conjecture on leading jets as a linear probe 
of the  collective nuclear glue. 
When reinterpreted in the reggeon field theory language,
our results correspond to a full resummation of multiple pomeron
exchanges enhanced by a large thickness of the target nucleus.
Such a resummation is interesting on its own, an extension
of these results beyond 
the dilute projectile and the Balitsky-Kovchegov approximations,
and the challenging inclusion of the pomeron loop effects (for the review 
see \cite{Trianta}) go beyond the scope of the present
communication.

The presentation of the main material is organized as follows. In
Sec. II we introduce the basic color dipole formalism, a definition
of the collective nuclear unintegrated glue and derive the
differential cross section of quasielastic scattering of quarks off
free nucleons and nuclei to the Born approximation. The color-dipole 
master formulas for the real and virtual
radiative corrections to the inclusive spectrum of leading quarks are  
presented in Sec. III. Our derivation of the virtual correction 
in Sec. III.B makes a manifest use of the unitarity relation.
We then proceed to the (nonlinear) $k_{\perp}$-factorization 
for the spectrum of leading quarks off the free nucleon (Sec. IV.A) and
color-triplet quark-nucleus (Sec. IV.B) collisions as well as
for the spectrum of  leading color-octet quarks (and gluons) in collisions with nuclei
(Sec. IV.C). These results are sufficient for
the quantitative treatment of the LPM effect at fixed transverse
momentum of leading jets at RHIC energies.
In Sec. IV.D we revisit 
the issue of unitarity and demonstrate the consistency of our 
radiative corrections with the leading quark number sum 
rule.  Inspired by the observations 
in Sec. II and Sec. IV, in Sec. V we focus on the  LL${1\over x}$ 
evolution properties of inclusive single particle spectra, which are
necessary for the extension of our technique to LHC energies.
We demonstrate, that to the  LL${1\over x}$ approximation, 
the nonlinear-evolving leading 
parton spectrum in parton-nucleus collisions proves to be 
linear $k_\perp$-factorizable in terms of the nonlinear-evolved
collective nuclear glue. 
In Sec. V.D we comment on the unified description of the LL${1\over x}$ 
evolution of the triplet-antitriplet and octet-octet ${\textsf S}$-matrices.
The subject of section V.E is a partial vindication of the BK
approximation - while it is not a closed-form equation and
is not really applicable to the free-nucleon target, for heavy
nuclei several iterations of the BK approximation will be a
useful working approximation in the range of energies from RHIC to LHC. 
This gives a practical procedure for
the quantitative evaluations of the LPM effect for leading jets
at LHC. 
In section V.F we comment on how our treatment
of the interplay between VRC and REC extends to the production 
of mid-rapidity jets, and on the possible interpretation
of our results from the effective reggeon field theory perspective.
In the Conclusions
we summarize our principal findings.
In the Appendix we comment on an
interesting observation that the nonlinear $k_\perp$-factorization
component of the LL${1\over x}$ nonlinear evolution 
gives a pure higher twist
contribution to the collective nuclear glue \cite{NonlinearBFKL}.


\section{Leading particle spectra to the Born approximation}


\subsection{Leading particles from the breakup of heavy quarkonium}

\begin{figure}[!t]
\begin{center}
\includegraphics[width = 7.5cm,height=3.5cm]{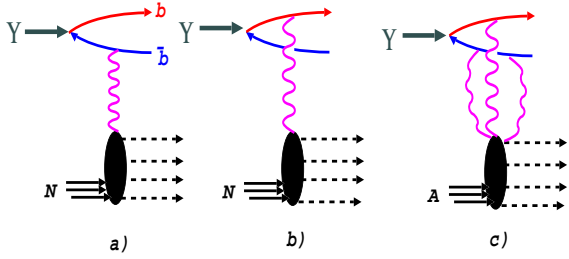}
\caption{(a,b) The Feynman diagrams for Upsilon breakup by
single-gluon exchange with the target nucleon and (c) an
archetype diagram of breakup of Upsilon by a multiple-gluon 
exchange with the target nucleus.}
\label{fig:UpsilonBreakup}
\end{center}
\end{figure}

To set up the framework, we start with a toy problem: the breakup
of the color-singlet heavy quarkonium (let it be $\Upsilon$) in a high energy 
collision with the free-nucleon
target $\Upsilon N \to b\bar{b} X$. Heavy quarkonium can 
well be approximated by its quark-antiquark Fock state
with the size $R_\Upsilon \gg 1/m_b$. The lowest
order pQCD subprocess is a breakup by Upsilon-gluon fusion,$\Upsilon g_t \to
b\bar{b}$,
where $g_t$ is the gluon exchanged in the $t$-channel. 
(Fig. \ref{fig:UpsilonBreakup}). The 
corresponding inclusive differential cross section,
obtained after summing over all excitations of the
target nucleon, is well known; it is found, 
for instance, in \cite{PionDijet} (Eqs. (4) and (5)):
\bea
{d\sigma(\Upsilon \to b\bar{b}) \over dz d^2\bp d^2\bDelta}=
{\alpha_S(\bp^2) \over 2\pi N_c}\cdot {{\cal F}(x,\bDelta^2) \over 
\Delta^4} 
\nonumber \\ \times
|\Psi_\Upsilon (z,\bp)-\Psi_\Upsilon (z,\bp-\bDelta)|^2.
\label{eq:2.A.1}
\eea
Here $\alpha_S(\bp^2)$ is the pQCD coupling, $N_c$ is the number of 
colors, $\Psi_\Upsilon (z,\bp)$ is the lightcone 
wave function of the quarkonium,
$\bp$ is the transverse momentum of the quark in the quarkonium, $z$
is the fraction of the lightcone momentum of the quarkonium carried 
by the quark, $\bDelta$ is the total transverse momentum of the 
$b\bar{b}$ pair,
\beq
{\cal F}(x,\kappa^2) = {\partial G(x, \kappa^2) \over \partial \log \kappa^2}
 \label{eq:2.A.2}
\eeq 
is the so-called  unintegrated gluon density in the target nucleon
as introduced in the BFKL approach \cite{KLF,BFKL}.
The Bjorken variable $x$ of the exchanged gluon
equals
\beq
x= {M_{b\bar{b}}^2 - M_\Upsilon^2 \over W_{\Upsilon p}^2},
 \label{eq:2.A.3}
\eeq 
where $ W_{\Upsilon p}$ is the cms energy in the $\Upsilon p$ system and
$M_{b\bar{b}}$ is the transverse mass of the excited quark-antiquark
pair,
\beq
\quad  M_{b\bar{b}}^2 =
{m_b^2+\bp^2 \over z} + {m_b^2+(\bp-\bDelta)^2 \over 1-z}. 
 \label{eq:2.A.4}
\eeq 
The transverse 
momentum $\bDelta$ of the $b\bar{b}$ pair  comes entirely from 
the exchanged gluon and the differential cross section (\ref{eq:2.A.1}) emerges as
a natural probe of the unintegrated glue ${\cal F}(x,\bDelta^2)$. 
We recall that for the free nucleon target ${\cal F}(x,\kappa^2)$ is
a solution of the LL${1\over x}$ BFKL evolution equation \cite{KLF,BFKL},
a detailed discussion of the connection between the integrated
glue $G(x,\kappa^2)$ of Eq. (\ref{eq:2.A.2}) and the gluon
densities of the leading order Dokshitzer-Gribov-Lipatov-Altarelli-Parisi
evolution equations \cite{DGLAP} is found in Refs.
\cite{Kfact,CollinsAPP,NZglue,INdiffglue,Smallx} and
need not be repeated here.

Now consider transverse momenta 
$\bp$ much larger than the typical momentum of the Fermi motion in
the quarkonium, $p_b \sim 1/R_\Upsilon$. The wave function  $\Psi(z,\bp)$ 
vanishes steeply beyond $p_b$, and 
the single-quark spectrum takes the convolution form
\bea
&&{d\sigma(\Upsilon \to b\bar{b}) \over dz d^2\bp }
= 
\int d^2\bkappa
{\alpha_S(\bp^2) \over 2\pi N_c}\cdot {{\cal F}(x,\bkappa^2) \over 
\bkappa^4} 
\nonumber \\ 
&&\times
|\Psi_\Upsilon (z,\bp)-\Psi_\Upsilon (z,\bp-\bkappa)|^2 
\nonumber\\
&&\approx \int d^2\bkappa
{\alpha_S(\bp^2) \over 2\pi N_c}\cdot {{\cal F}(x,\bkappa^2) \over 
\bkappa^4} |\Psi_\Upsilon (z,\bp-\bkappa)|^2 \nonumber\\
&&=  \int d^2\bkappa d^2\bp_b \delta(\bp -\bp_b -\bkappa)\nonumber\\
&&\times
 {dq_{b/\Upsilon}(z,\bp_b) \over  d^2\bp_b}\cdot  {d\sigma(b\to b')\over 
d^2\bkappa}.
\label{eq:2.A.5}
\eea
Here 
\beq 
{dq_{b/\Upsilon}(z,\bp_b) \over d^2\bp_b}=  |\Psi_\Upsilon (z,\bp_b)|^2
\label{eq:2.A.6}
\eeq
is the  momentum distribution of quarks in the quarkonium
and 
\beq
{d\sigma(b\to b')\over 
d^2\bkappa}= {\alpha_S(\bp^2) \over 2\pi N_c}\cdot {{\cal F}(x,\bkappa^2) \over 
\bkappa^4} 
\label{eq:2.A.7}
\eeq
must be interpreted as a differential cross section of the 
valence quark quasielastic scattering $bN\to b'X$.
Furthermore, the longitudinal momentum distribution of leading quarks
is given by precisely the transverse momentum-dependent $z$-distribution 
in the quarkonium -- $z$ is the Feynman variable of the observed 
quark. As such, the convolution (\ref{eq:2.A.5}) can be 
regarded as the archetype $k_\perp$-factorization for the production 
of hard leading quark jets in high-energy hadron-nucleon collisions.
The adjective `hard' refers to the jet momenta harder than
the intrinsic momentum of quarks in the quarkonium: 
$\bp^2 \gg \langle p_b^2\rangle$.
Now we proceed to the $k_\perp$-factorization
properties of the quasielastic scattering of partons -- quarks and
gluons -- off free nucleons and heavy nuclei.


\subsection{The free-nucleon target: the dipole cross section, unintegrated glue and
quasielastic $aN$ scattering}

\begin{figure}[!t]
\begin{center}
\includegraphics[width = 7.5cm]{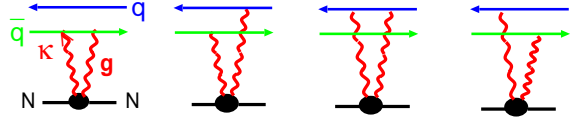}
\caption{The four Feynman diagrams for the quark-antiquark
dipole-nucleon interaction by the two-gluon pomeron exchange
in the $t$-channel.}
\label{fig:SigmaDipoleFeynman}
\end{center}
\end{figure}

We start with the dipole cross section for the $a\bar{a}$ color dipole.
It is described by two-gluon exchange in the $t$-channel
(Fig. \ref{fig:SigmaDipoleFeynman}). The $\textsf{S}$-matrices of 
the $aN$ and $\bar{a}N$ interaction at impact parameter $\bb_a(\bb_{\bar{a}})$ 
equal, respectively \cite{Nonlinear}, 
\bea 
\textsf{S}_a(\bb_a) & =& \openone
+ i\textsf{T}_a^\alpha \hat{V}_\alpha\chi(\bb_a)- {1\over 2}
\textsf{T}_a^\alpha\textsf{T}_a^\alpha \chi^2(\bb_a)\, , 
\nonumber\\
\textsf{S}_a^\dagger(\bb_{\bar{a}}) & =& \openone
- i\textsf{T}_a^\alpha \hat{V}_\alpha \chi(\bb_{\bar{a}}) - {1\over 2}
\textsf{T}_a^\alpha\textsf{T}_a^\alpha \chi^2(\bb_{\bar{a}}),
\nonumber\\
\label{eq:2.B.1}
\eea
were $\textsf{T}_a^\alpha \hat{V}_\alpha\chi(\bb)$ is the eikonal 
operator for the $aN$ single-gluon exchange interaction. 
In the realm of the Standard Model QCD, quarks belong to the color-triplet
(fundamental) representation of $SU(N_c)$, but much light on the
small-$x$ evolution properties of the nuclear spectra is shed by
considering quarks also in the octet (adjoint) representation of
$SU(N_c)$. One can also think of leading gluons in interactions of
glueballs. The expansion
(\ref{eq:2.B.1}) to the
considered pQCD approximation satisfies the unitarity condition
\bea 
\textsf{S}_a(\bb_a)\textsf{S}_a^\dagger(\bb_a)=\openone.
\label{eq:2.B.2}
\eea
The spin of partons is
unimportant in view of the $s$-channel helicity conservation
in high energy QCD. The 
important observation is that $\textsf{S}_a^\dagger$ equals the
$\textsf{S}$-matrix for interaction of the antiparticle $\bar{a}$
\cite{SlavaPositronium,NPZcharm}. The vertex 
operator $\hat{V}_\alpha$ for excitation
of the nucleon $g^\alpha N \to N^*_\alpha$ into a color octet state 
$N^*_\alpha$ is so
normalized that after the application of closure over
the final state excitations $N^*$ the vertex $g^\alpha g^\beta
NN$ equals $\bra{ N} \hat{V}_\alpha^\dagger \hat{V}_\beta \ket{ N}=
\delta_{\alpha\beta}$. 
The second order terms in (\ref{eq:2.B.1})
do already use this normalization.

The $\textsf{S}$-matrix of the
$(a\bar{a})$-nucleon interaction equals
\bea
\textsf{S}_{a\bar{a}}(\bb_a,\bb_{\bar{a}}) =
{\bra{N}{\rm Tr}[\textsf{S}_a(\bb_a)\textsf{S}_a^\dagger(\bb_{\bar{a}})] 
\ket{N} \over \bra{N} {\rm Tr} \openone  \ket{N} }
\nonumber\\
=
1 - {1\over 2} C_a[\chi(\bb_a)-\chi(\bb_{\bar{a}})]^2\,.
\label{eq:2.B.3}
\eea
where $C_a$ is the quadratic Casimir operator.
The corresponding profile function is 
$\Gamma_{a\bar{a}}(\bb_a,\bb_{\bar{a}})= 1 - \textsf{S}_{a\bar{a}}(\bb_a,\bb_{\bar{a}})$,
and the dipole cross section for interaction of the color-singlet
$a\bar{a}$ dipole $\br=\bb_a-\bb_{\bar{a}}$  with the free
nucleon is obtained
upon the integration over the overall impact parameter 
\bea
\sigma_{a\bar{a}}(x,\br) &=& 2\int d^2\bb \, 
\Gamma_{a\bar{a}}(\bb,\bb-\br) 
\nonumber \\
&=&
C_a \int
d^2\bb [\chi(\bb)-\chi(\bb-\br)]^2 
\nonumber \\
&=& {C_a\over C_F} \sigma(x,\br)\, , 
\label{eq:2.B.4} 
\eea
where $C_F=(N_c^2-1)/2N_c$ is the quark Casimir and 
$\sigma(x, \br)$ is the dipole cross section 
for the $q\bar{q}$-dipole.
It is related to the gluon density in the target by
the $k_{\perp}$-factorization formula \cite{NZglue,NZ94}
\bea 
\sigma(x,\br) &=& \int d^2\bkappa f (x,\bkappa)
[1-\exp(i\bkappa\br)]\,,
\label{eq:2.B.5} 
\eea
where
\bea
f (x,\bkappa)&=& {4\pi \alpha_S \over N_c}\cdot {1\over \kappa^4} \cdot {\cal
F}(x,\bkappa^2).
\label{eq:2.B.6} 
\eea

The leading Log${1\over x}$ evolution of the dipole
cross section is governed by the color-dipole BFKL evolution
\cite{NZ94,NZZBFKL}, the same evolution for the
unintegrated gluon density is governed by the familiar
momentum-space BFKL equation \cite{KLF,BFKL}.
For very large dipoles 
\bea
\sigma_{a\bar{a},0}(x)&=&
C_a \int
d^2\bb [\chi^2(\bb)+\chi^2(\bb-\br)]
\nonumber \\
&=&{C_a \over C_F} \int d^2\bkappa f (x,\bkappa) 
\nonumber\\
& =& 2C_a\int
d^2\bb {\bra{N}{\rm Tr}\{ \openone-\textsf{S}_a(\bb)\}
\ket{N} \over \bra{N} {\rm Tr}  \openone \ket{N} }
\nonumber \\
&=& 2\sigma_a(x) = {C_a
\over C_F} \sigma_0(x),
 \label{eq:2.B.7}
\eea 
where $\sigma_0(x) $ is the cross section for large $q\bar{q}$ dipoles.

Now we turn to the master formula for the inclusive 
differential cross section 
of quasielastic scattering  $aN \to a'N^*$.
Only
the final states $N^*\neq N$ are excited by
the $t$-channel gluon exchange and 
\begin{widetext}
\bea
{d{\sigma}_{Qel} \over d^2\bp} &=&
{1\over (2\pi)^2}\int d^2\bb d^2\bb'\exp[-i\bp(\bb-\bb')]
\sum_{N^*\neq N} {\bra{N}{\rm Tr}\Big\{[\openone -\textsf{S}_a(\bb)]\ket{N^*}
\bra{N^*}[\openone -\textsf{S}_a^\dagger(\bb')] \Big\}
\ket{N} \over \bra{N}  {\rm Tr} \openone  \ket{N}}. \nonumber \\
\label{eq:2.B.8}
\eea
\end{widetext}
After the application of closure 
\beq
\sum_{N^*\neq N}\ket{N^*} \bra{N^*}+ \ket{N} \bra{N}=1,
\label{eq:2.B.9}
\eeq
and upon simple algebra one finds
\bea
{d{\sigma}_{Qel} (x)\over d^2\bp}&=& {1\over 2(2\pi)^2}\int d^2\bc
\exp[-i\bp\bc]
\nonumber\\
&\times&
[\sigma_{a\bar{a},0}(x)-\sigma_{a\bar{a}}(x,\bc)]\nonumber\\
&=&
{C_a\over 2C_F}f(x,\bp),
\label{eq:2.B.10}
\eea
where $\bp$ is the transverse momentum of the scattered quark,
cf. with Eq. (\ref{eq:2.A.7}). 
 To this Born approximation,
quasielastic scattering off the nucleon exhausts the total cross
section, the pure elastic cross section is of higher order in pQCD 
perturbation theory - it starts with four gluons in the $t$-channel.
 Quasielastic  $qN$ scattering 
emerges as  a direct probe of the unintegrated gluon structure function
of the target nucleon -- we shall show this to hold beyond
the Born approximation. 
The above explicit calculation of the emerging single-nucleon
matrix elements is straightforward and hereafter, unless that might 
cause a confusion, it will be suppressed.


\subsection{Quasielastic $qA$ scattering and
collective nuclear glue}
Here we need to specify the boundary condition of the small-x evolution.
In the interaction with heavy nuclei the coherent nuclear 
effects evolve, and the concept of the collective
unintegrated  nuclear glue becomes applicable, for partons
with
\beq
x \lsim x_A = {1\over 2R_A m_N}.
\label{eq:2.C.1}
\eeq
In the Breit frame, this coherence condition corresponds to
the spatial overlap and fusion of partons belonging to different nucleons
at the same impact parameter in the Lorentz-contracted
ultrarelativistic nucleus \cite{NZfusion}.
At $x\gsim x_A$ coherent nuclear effects are absent and the
impulse approximation holds. The small-$x$ evolution down to
$x_A$, the span
of which could be substantial for the parametrically small
$x_A$ for very heavy nuclei, must be performed at the free
nucleon level. The evaluation of coherent nuclear effects at
$x < x_A$ must start with the boundary conditions set by
the free-nucleon quantities evaluated at $x_0=x_A$. We refer
to this boundary condition as the nuclear Born approximation.

Interaction of the generic  $n$-parton system with the 
target nucleon is described by the coupled-channel cross
section operator $\hat{\Sigma}^{(n)}$. In the regime of coherent 
interaction with the heavy nucleus, $x\lsim x_A$, the corresponding
$\textsf{S}$-matrix operator is given by the Glauber-Gribov 
formula \cite{Glauber,Gribov}
\bea 
\textsf{S}[\bb,\hat{\Sigma}^{(n)}]=\exp[-{1\over 2}\hat{\Sigma}^{(n)}T(\bb)]\, , 
\label{eq:2.C.2} 
\eea 
where
\bea
T(\bb)= \ds \int_{-\infty}^\infty dr_z \, n_A(\bb,r_z)
\label{eq:2.C.3} 
\eea 
is the optical thickness of the nucleus. The nuclear density $n_A(\bb,r_z)$ is
normalized according to $\ds \int d^3\vec{r} \, n_A(\bb,r_z) = \int
d^2\bb\,  T(\bb) = A$, where $A$ is the nuclear mass number. 
The principal assumption behind (\ref{eq:2.C.2}) is that the nucleus
is {\it a dilute uncorrelated gas of color-singlet nucleons}. Also, color dipoles 
are  much smaller than the size of the nucleus. All
the nuclear cross sections are defined per unit area in the impact parameter
space. 

For the transformation from the color-dipole to the momentum
representation, we define the collective nuclear unintegrated gluon density 
per unit area in the impact parameter plane, $\phi(\bb,x_0,\bkappa)$,
in terms of the Born-approximation nuclear profile function for the 
triplet-antitriplet $q\bar{q}$ dipole
\cite{NSSdijet,NSSdijetJETPLett,Nonlinear,NonlinearJETPLett}:
\bea 
\Gamma[\bb,\sigma(x_0,\br)] &=& 1-{\textsf S}[\bb;\sigma(x_0,\br)]
\nonumber\\
&=& 1-\exp\left[-{1\over 2}\sigma(x_0,\br)T(\bb)\right]
\nonumber \\
&\equiv& \int d^2\bkappa
\phi(\bb,x_0,\bkappa) \Big[1 - \exp(i\bkappa\br) \Big] \, .
\nonumber \\
\label{eq:2.C.4} 
\eea
The expansion for $\phi(\bb,x_0,\bkappa)$ in terms of the collective
glue for overlapping nucleons in the Lorentz-contracted 
ultrarelativistic nucleus, and its nuclear shadowing and
antishadowing properties, are found in \cite{Nonlinear,SingleJet}. 
The utility of $\phi(\bb,x_0,\bkappa)$ stems from the 
observation that the driving term of small-$x$ nuclear structure 
functions, the amplitude of coherent diffractive production 
of dijets off nuclei and the single-quark
spectrum from the $\gamma^* \to q\bar{q}$ excitation off a nucleus 
all take the familiar linear $k_{\perp}$-factorization form in terms of
$\phi(\bb,x_0,\bkappa)$. 
Still more convenient, and more universal, 
would be a definition in terms of the 
color-dipole $\textsf{S}$ matrix,
\bea
\Phi(\bb,x_0,\bkappa)&=& \int {d^2\br  \over (2 \pi)^2}\textsf{S}[\bb,\sigma(x_0,\br)]
\exp(-i\bkappa \cdot \br) \nonumber\\
&=&
\textsf{S}[\bb,\sigma_{0}(x_0)]\delta^{(2)}(\bkappa)
+\phi(\bb,x_0,\bkappa),
\nonumber \\
\label{eq:2.C.5}
\eea
which by its definition satisfies the sum rule
\beq
\int d^2\bkappa \Phi(\bb,x_0,\bkappa) = 1.
\label{eq:2.C.6}
\eeq

Now we proceed to the differential cross section 
of quasielastic scattering  $aA \to aA^*$.  Heavy nuclei
are strongly absorbing targets and the optical nuclear thickness 
$T(\bb)$ is a new large parameter in the problem. 
Because of enhanced multiple 
$t$-channel gluon exchanges the pure elastic scattering
is no longer suppressed and final states $A^*$ include the 
ground state of the nucleus. The calculation of the total $\bp$
spectrum of quarks in $qA$ scattering proceeds as
follows. The master formula is
\begin{widetext}
\bea
{d{\sigma}_{el+Qel}^B \over d^2\bp d^2\bb}& =&{1\over (2\pi)^2}
\int d^2\bc\exp(-i\bp\cdot\bc)
\sum_{A^*} {\bra{A}{\rm Tr}\Big\{[\openone-\textsf{S}_a(\bb)]\ket{A^*}
\bra{A^*}[\openone-\textsf{S}_a^\dagger(\bb-\bc)] \Big \}
\ket{A} \over \bra{A} {\rm Tr} \openone  \ket{A} }.
\nonumber \\
\label{eq:2.C.7}
\eea
\end{widetext}
The nuclear $\textsf{S}$-matrix is a product of the 
free-nucleon $\textsf{S}$-matrices.
The technique of the calculation of nuclear matrix elements in the standard
picture of a nucleus as a dilute uncorrelated gas of color-singlet
nucleons is found in \cite{Glauber,Nonlinear}
and need not be repeated here.
We simply cite the results,
\bea
&&{\bra{A}{\rm Tr}\{\textsf{S}_a(\bb)\}\ket{A} \over \bra{A}  
{\rm Tr}\openone \ket{A}  }= \textsf{S}[\bb,\sigma_a(x_0)],\nonumber\\
&&
 {\bra{A}{\rm Tr}\{\textsf{S}_a(\bb)\textsf{S}_a^\dagger(\bb-\bc)\}
\ket{A} \over \bra{A} {\rm Tr} \openone \ket{A} }=  
\textsf{S}[\bb,\sigma_{a\bar{a}}(x_0,\bc)].
\nonumber \\
\label{eq:2.C.8}
\eea
Now we apply (\ref{eq:2.C.5}) and notice that 
\beq
\textsf{S}[\bb,\sigma_{a\bar{a},0}(x_0)]= 
\textsf{S}^2[\bb,\sigma_{a}(x_0)],
\label{eq:2.C.9}
\eeq
which leads to
\bea
{d{\sigma}_{el+Qel}^B(x_0) \over d^2\bp d^2\bb}=
\{ 1-\textsf{S}[\bb,\sigma_{a}(x_0)]\}^2
\delta^{(2)}(\bp) 
\nonumber \\
+ \phi_{a\bar{a}}(\bb,x_0,\bp).
\label{eq:2.C.10}
\eea
The first term describes the elastic scattering, here the 
delta-function $\delta^{(2)}(\bp)$ is a simple
working approximation to the nuclear diffraction peak. The
second term in the $\bp$ spectrum (\ref{eq:2.C.10}) describes 
the Born (B) approximation for the inclusive quasielastic  scattering $aA\to aX$,
\beq
{d{\sigma}_{Qel}^B(x_0) \over d^2\bp d^2\bb}=
 \phi_{a\bar{a}}(\bb,x_0,\bp),
\label{eq:2.C.11}
\eeq
where $\phi_{a\bar{a}}(\bb,x_0,\bp)$ is defined in terms of   
\beq
\textsf{S}[\bb,\sigma_{a\bar{a}}(x_0,\br)]=
\textsf{S}[\bb,{C_a\over C_F} \sigma(x_0,\br)].
\label{eq:2.C.12}
\eeq
Note, that the color representation dependence of the
dipole cross section (\ref{eq:2.B.4}) entails a variety of the so-defined 
collective nuclear glue $\Phi_{a\bar{a}}(\bb,x_0,\bkappa)$ -- the
nuclear glue is a 
density matrix in color space.
The major point is that to the considered Born approximation
 quasielastic scattering is linear $k_\perp$-factorizable
in terms of the collective glue  $\phi_{a\bar{a}}(\bb,x_0,\bp)$.

Upon the
integration of nuclear Born cross sections over all transverse momenta 
\bea 
\sigma_{el} &=& \int d^2\bb\{ 1-\textsf{S}[\bb,\sigma_{a}(x_0)]\}^2 ,\nonumber\\
\sigma_{in} &=& \sigma_{Qel} = \int d^2\bb\{ 1-\textsf{S}^2[\bb,\sigma_{a}(x_0)]\},\\
\sigma_{tot}&=& \sigma_{el}+\sigma_{in} = 
2\int d^2\bb\{ 1-\textsf{S}[\bb,\sigma_{a}(x_0)]\},
\nonumber \\
\label{eq:2.C.13}
\eea
which are the standard Glauber formulas for the the projectile with $\sigma_{tot}^{aN}=
\sigma_a(x_0)$ \cite{Glauber}.


\section{Master formulas for the real and virtual radiative corrections
to inclusive single leading jet production}


\subsection{Real production $a \to ag$}

The leading order pQCD radiative correction to the spectrum of 
leading particles in $aN(A)$ scattering, $a=q,g$,
comes from the radiation of gluons, $aN(A) \to ag N^*(A^*)$.
To the lowest order in pQCD, 
the generic partonic subprocess is $ag_t \to bc$. 
From the laboratory frame standpoint, it can
be viewed as an excitation of the perturbative $|bc\rangle$ Fock state
of the physical projectile $|a\rangle$ by one-gluon exchange with the
target nucleon. In the case of a nuclear target one has to deal with
multiple gluon exchanges which are enhanced by a large thickness of the
target nucleus. Here the 
frozen impact parameter approximation holds, 
if the coherence length $l_c$ is larger than the diameter of the nucleus $2R_A$,
\beq
l_c ={ 2E_{a} \over (Q^*)^2 + M_{\perp}^2} = {1\over x m_N}> 2R_A\,,
\label{eq:3.A.1}
\eeq
where
\beq
M_{\perp}^2 = {\bp_b^2 +m_b^2\over z_b}+{\bp_c^2+m_c\over z_c}
\label{eq:3.A.2}
\eeq
is the transverse mass squared of the $bc$ state,
$\bp_{b,c}$ and $z_{b,c}$ are the transverse momenta and
fractions of the the incident parton's momentum carried by the
outgoing partons ($z_b+z_c=1$). Notice an equivalence 
between the coherency condition (\ref{eq:3.A.1}) and the parton fusion
condition (\ref{eq:2.C.1}).
The virtuality of the incident
parton $a$ equals $(Q^*)^2=\bk_a^2$, where $\bk_a$ is the
transverse momentum of the parton $a$ in the incident proton.

\begin{figure}[!t]
\begin{center}
\includegraphics[width = 7.0cm, angle = 0]{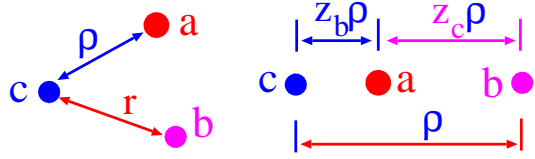}
\caption{The color dipole structure (lhs) of the generic 3-parton state and 
(rhs) of the 3-parton state entering the color-dipole description
of fragmentation $a\to bc$ with 
formation of the $bc$ dipole of size $\brho$.}
\label{fig:ABCdipole}
\end{center}
\end{figure}

The target frame rapidity structure of the considered $a\to bc$ 
excitation 
is shown in fig.~1. The beam parton has a rapidity $\eta_a \gsim \eta_A=\log{1/x_A}$, 
the final state partons too
have rapidities  $\eta_{b,c} \gsim \eta_A$. In this paper we focus
on the lowest order radiative corrections without 
production of more secondary partons in the rapidity span
 between $\eta_a$ ($\eta_{q'}$)  and $\eta_c$ ($\eta_g$).

To the lowest
order in the perturbative transition $a\to bc$ the Fock state expansion for
the physical state $|a_{phys}\rangle$ reads
\beq
 \ket{a_{phys}} = \sqrt{Z_a} \ket{a}_0 + \Psi(z_b,\brho) \ket{bc}_0
\label{eq:3.A.3}
\eeq
where $\Psi(z_b,\brho)$ is the probability amplitude to find the $bc$ system
with the separation $\brho$ in the two-dimensional impact parameter space,
the subscript $"0"$ refers to bare partons. The explicit dependence
of the wave functions on the virtuality $(Q^*)^2$ and their relation to
the parton splitting functions is found in \cite{SingleJet}.
Here $Z_a$ is the wave function renormalization, the perturbative
coupling of the $a\to bc$ transition is reabsorbed into the lightcone
wave function $\Psi(z_b,\brho)$. The normalization (completeness)
 condition reads
\beq
1=Z_a  + \int |\Psi(z_b,\brho)|^2 \,  d^2\brho dz_b
\label{eq:3.A.4}
\eeq

If $\bb_a=\bb$ is the impact parameter of the projectile $a$, then
\beq
\bb_{b}=\bb+z_c\brho, \quad\quad \bb_{c}=\bb-z_b\brho\, ,
\label{eq:3.A.5}
\eeq
see Fig.~\ref{fig:ABCdipole}. 
Below we cite all
the spectra in the $a$-target collision frame. We shall speak of the 
produced parton $b$ - or jet originating from this parton -
as the {\it leading} one if it carries a large fraction of
the beam lightcone momentum, $z_b\sim 1$. The Feynman variable $x_F$
spectra in $pN,pA$ collisions  are obtained from the $z_b$ spectra 
in $qN, qA$ collisions by an obvious $k_\perp$-factorization
convolution with the beam quark densities, see below 
Eq. (\ref{eq:4.A.12}).

By the conservation of
impact parameters, the action of the $S$-matrix on $\ket{a_{phys}}$
takes a simple form
\bea
\textsf{S}\ket{a_{phys}} &=&
S_a(\bb) \sqrt{Z_a} \ket{a}_0 \nonumber\\
&+&
S_b(\bb_b) S_c(\bb_c)\Psi(z_b,\brho) \ket{bc}_0 
\nonumber \\
&=& S_a(\bb) \ket{a_{phys}} \nonumber\\
&+&  [ S_b(\bb_b) S_c(\bb_c) - S_a(\bb) ]
\Psi(z_b,\brho) \ket{bc} .  \nonumber\\
\label{eq:3.A.6} \eea 
In the last line we
explicitly decomposed the final state into the elastically
scattered $\ket{a_{phys}}$ and the excited state $\ket{bc}_{0}$.
The two terms in the latter describe a scattering on the target of
the $bc$ system formed way in front of the target and the
transition $a\to bc$ after the interaction of the state
$\ket{a}_0$ with the target, as illustrated in
Fig. \ref{fig:single-jet_a_to_bc}.
 The contribution from transitions
$a\to bc$ inside the target nucleus vanishes in the high-energy
limit of $x\lsim x_A$.

\begin{widetext}
The probability amplitude for the two-parton spectrum is given by the
Fourier transform
\begin{figure}[!t]
\begin{center}
\includegraphics[width = 4.5cm, angle = 270]{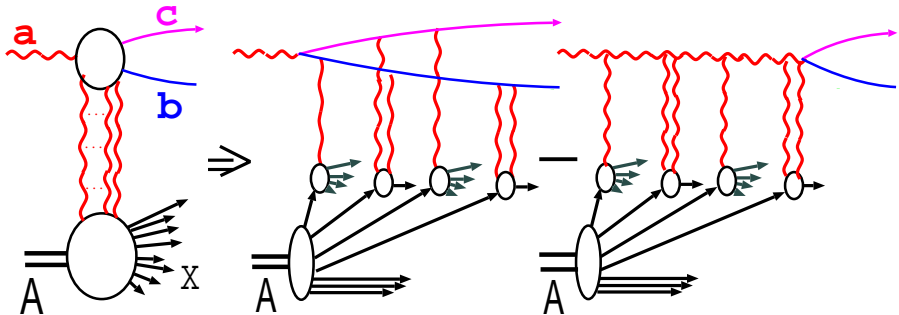}
\caption{Typical contribution to the excitation amplitude for $a A \to bc  X$,
with multiple color excitations of the nucleus. 
The amplitude receives contributions from processes that involve interactions 
with the nucleus after and before the virtual 
decay $a\to bc$ which interfere destructively.}
\label{fig:single-jet_a_to_bc}
\end{center}
\end{figure}
\beq 
\int d^2\bb_b d^2\bb_c \exp[-i(\bp_b\bb_b +
\bp_c\bb_c)][ \textsf{S}_b(\bb_b) \textsf{S}_c(\bb_c) - \textsf{S}_a(\bb) ] 
\Psi(z_b,\brho).
\label{eq:3.A.7} 
\eeq 
The differential cross section is proportional
to the modulus squared of (\ref{eq:3.A.7}) 
\bea
&&{d \sigma (a^* \to b(\bp_b) c(\bp_c)) \over dz d^2\bp_b d^2\bp_c } = 
{1 \over (2 \pi)^4} \int
d^2\bb_b d^2\bb_c d^2\bb'_b
 d^2\bb'_c\exp[i \bp_b
(\bb_b -\bb'_b) + i
\bp_c(\bb_c
-\bb_c')] \nonumber \\
&&\times \Psi_{bc}(z_b,\bb_b -
\bb_c) \Psi^*_{bc}(z_b,\bb'_b-\bb'_c)
\Big\{
\textsf{S}^{(4)}_{\bar{b}\bar{c} c b}(\bb_b',\bb_c',\bb_b,\bb_c) 
+ \textsf{S}^{(2)}_{\bar{a}a}(\bb',\bb) -
\textsf{S}^{(3)}_{\bar{b}\bar{c}a}(\bb,\bb_b',\bb_c')
- \textsf{S}^{(3)}_{\bar{a}bc}(\bb',\bb_b,\bb_c) \Big\}.\nonumber\\
\label{eq:3.A.8} 
\eea
\begin{figure}[!t]
\begin{center}
\includegraphics[width = 4.5cm,angle=270]{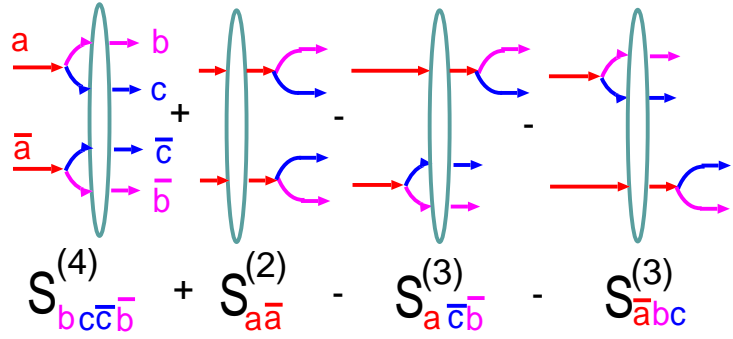}
\caption{The $\textsf{S}$-matrix structure of the two-body density
matrix for excitation $a\to bc$.}
\label{fig:SingleJetDensityMatrix}
\end{center}
\end{figure}
\end{widetext}
where, see  Fig.~\ref{fig:SingleJetDensityMatrix} for an illustration,
\bea
\textsf{S}^{(2)}_{\bar{a}a}(\bb',\bb)&=& 
\textsf{S}_a^{\dagger}(\bb')\textsf{S}_a(\bb),
\nonumber \\ \textsf{S}^{(3)}_{\bar{a}bc}(\bb',\bb_b,\bb_c) &=&
\textsf{S}_a^{\dagger}(\bb')\textsf{S}_b(\bb_b) \textsf{S}_c(\bb_c), \nonumber \\
\textsf{S}^{(3)}_{\bar{b}\bar{c}a}(\bb,\bb_b',\bb_c') &=&
\textsf{S}_b^{\dagger}(\bb_b')\textsf{S}_c^{\dagger}(\bb_c') 
\textsf{S}_a(\bb), \nonumber \\
\textsf{S}^{(4)}_{\bar{b}\bar{c} c b}(\bb_b',\bb_c',\bb_b,\bb_c) &=&
\textsf{S}_b^{\dagger}(\bb_b')\textsf{S}_c^{\dagger}(\bb_c') 
\textsf{S}_c(\bb_c)\textsf{S}_b(\bb_b) \,.\nonumber \\
\label{eq:3.A.9}
\eea

Here the proper averaging over polarizations 
and color states of the initial parton and the summation over polarizations 
and color states of final state partons are  understood.
Next we approximate the heavy nucleus by a dilute gas of colorless 
nucleons and follow Glauber's technique of summing over the final states
of the target nucleus applying the closure 
relation \cite{Glauber}. Then the nuclear matrix element of products of
single-parton $\textsf{S}$-matrices would boil down to the Glauber-Gribov
formula (\ref{eq:2.C.2}):
$\bra{A}\textsf{S}^{(n)} \ket{A} = \textsf{S}[\bb;\hat{\Sigma}^{(n)}]$,
where $\hat{\Sigma}^{(n)}$ is the relevant $n$-parton cross section operator, for the
details see Refs. \cite{Nonlinear,QuarkGluonDijet,GluonGluonDijet}. 
All the multiparton states -- $\bar{a}a$, 
$\bar{a}bc$, $\bar{b}\bar{c}a$, $\bar{b}\bar{c}cb$ -- are in an overall 
color-singlet state and the excitation cross sections are 
infrared-finite despite the nonvanishing net color charge
of the projectile parton $a$ \cite{NPZcharm,SingleJet}.


Integration over the transverse momentum $\bp_c$ of the jet $c$
gives
\bea
\bb_c &= &\bb_c' \,, \quad
\bb - \bb' =
z_b(\brho-\brho')\, ,
\nonumber \\
\quad \bb_b-\bb_b'&=& \brho-\brho'\, , \quad
\bb'-\bb_c=z_b\brho'\, . 
\label{eq:3.A.11} 
\eea

Then the unitarity relation 
\beq
S_c^{\dagger}(\bb_c) S_c(\bb_c)=1
\label{eq:3.A.12}
\eeq
leads to a fundamental simplification:
\bea
S^{(4)}_{\bar{b}\bar{c} c b} (\bb_b',\bb_c',\bb_b,\bb_c) &=&
S_b^{\dagger}(\bb_b')S_c^{\dagger}(\bb_c) S_c(\bb_c)S_b(\bb_b)
\nonumber \\
&=&
S_b^{\dagger}(\bb_b')S_b(\bb_b)
\nonumber \\
&=&S^{(2)}_{\bar{b}b}(\bb_b',\bb_b) \
, 
\label{eq:3.A.13}
\eea
i.e., the effect of interactions of the spectator parton $c$ vanishes 
upon the summation over all its color states and 
integration over all its transverse
momenta (\cite{NPZcharm}, see also a discussion in \cite{SingleJet}) 
the only trace of the observed parton $b$ having 
been produced in the fragmentation
$a\to bc$ is in the density matrix $\Psi(z_b,\bb_b -
\bb_c) \Psi^*(z_b,\bb'_b-\bb'_c)$ which defines the transverse
 momentum distribution
of the parton $b$ in the beam parton $a$ and the partition 
$z_b \,,\,z_c= (1-z_b)$, of
the longitudinal momentum between the final state partons.
The applications of the master formula (\ref{eq:3.A.8}) to the calculation 
of the real emission spectra for the free-nucleon and nuclear targets 
are found in \cite{SingleJet}.


\subsection{Derivation of  the virtual radiative correction
from the unitarity relation}

The relevance of the unitarity has already been explained in the Introduction.
It is especially obvious in $qN(A)$ scattering. The opening of the new 
channel $q\to qg$ renormalizes, via unitarity, the Born cross section of the
lower order quasielastic scattering $qN(A)\to q'N^*(A^*)$.  
The VRC is precisely that unitarity driven renormalization.
The real emission and the
virtual correction must so conspire, that the total multiplicity of
leading quarks
in the final state is exactly unity. We show here two derivations which
have already been used in the related works \cite{NPZcharm,NZ94,NZZBFKL}.

The first approach is a direct perturbative 
solution of the unitarity relation
(\cite{NPZcharm}, for an extensive use in the theory
of the LPM effect see \cite{SlavaLPM}).
We define the radiatively corrected $\textsf{S}$-matrix of elastic scattering, 
\beq
\textsf{S}_{el}(\bb,x)= \textsf{S}_a(\bb,x_0)+\delta \textsf{S}_a(\bb,x),
\label{eq:3.B.1} 
\eeq
and the excitation (real emission) operator 
\beq
\textsf{S}_{ex}(\bb,x)= 
\textsf{S}_b(\bb_b,x_0) \textsf{S}_c(\bb_c,x_0) - \textsf{S}_a(\bb,x_0).
\nonumber\\  
\label{eq:3.B.2} 
\eeq
Then the 
unitarity relation would read 
\bea
&&\textsf{S}(\bb,x)\textsf{S}^\dagger(\bb,x) = 
\textsf{S}_{el}(\bb,x) \textsf{S}_{el}^\dagger(\bb,x)
\nonumber \\
&&+ \int {\cal D}_{bc}\Psi^*_{bc}\Psi_{bc} 
\textsf{S}_{ex}(\bb,x)\textsf{S}_{ex}^\dagger(\bb,x)=1,
\label{eq:3.B.3}
\eea
where we explicitly separated the contributions from the elastic
and real emission channels, 
${\cal D}_{bc}$ is the short hand notation for the integral over the $bc$ dipole
parameters, ${\cal D}_{bc} = d^2\brho dz_c$. If the boundary condition for
the small-$x$ evolution is defined at $x_0$, then the integration over
$dz_c$ extends over 
the interval $[x/x_0,1]$. We recall that  
$\Psi^*_{bc}\Psi_{bc} \propto \alpha_S$
and the real emission contribution to (\ref{eq:3.B.3}) must be regarded 
as the pQCD perturbation. Making use of the explicit form 
of $\textsf{S}_{ex}(\bb,x)$, 
to the linear order in $\delta \textsf{S}_a(\bb,x)$ we find
\begin{widetext}
\bea
\delta \textsf{S}_a(\bb,x) \textsf{S}_a^\dagger(\bb,x_0) +  \textsf{S}_a(\bb,x_0)
\delta \textsf{S}_a^\dagger(\bb,x)
&=&  \int{\cal D}_{bc} \Psi^*_{bc}\Psi_{bc} 
[ \textsf{S}_b(\bb_b,x_0) 
\textsf{S}_c(\bb_c,x_0)\textsf{S}_a^\dagger(\bb,x_0)-1]
\nonumber\\
&+&  \int {\cal D}_{bc}\Psi^*_{bc}\Psi_{bc} 
[\textsf{S}_a(\bb_a,x_0) 
\textsf{S}_c^\dagger(\bb_c,x_0)\textsf{S}_b^\dagger(\bb_b,x_0)-1].
\label{eq:3.B.4}
\eea
Evidently, this equation can be split into 
\bea
\delta \textsf{S}_a(\bb,x) \textsf{S}_a^\dagger(\bb,x_0)  =  
\int {\cal D}_{bc}\Psi^*_{bc}\Psi_{bc} 
[ \textsf{S}_b(\bb_b,x_0) 
\textsf{S}_c(\bb_c,x_0)\textsf{S}_a^\dagger(\bb,x_0)-1]
\label{eq:3.B.5}
\eea
and its complex conjugate. Multiplication of (\ref{eq:3.B.5}) by 
$\textsf{S}_a(\bb,x_0)$ from the right and the application of the unitarity
relation (\ref{eq:2.B.2}) gives the solution 
\beq
\delta \textsf{S}_a(\bb,x) = \int {\cal D}_{bc}\Psi^*_{bc}\Psi_{bc}
[ \textsf{S}_b(\bb_b,x_0) \textsf{S}_c(\bb_c,x_0)-\textsf{S}_a(\bb,x_0)]
\label{eq:3.B.6}
\eeq

The second approach is a direct calculation 
\bea
\textsf{S}_{el}(\bb,x)&=& \bra{a_{phys}} \textsf{S}\ket{a_{phys}} =
 Z_a  \textsf{S}_a(\bb,x_0)+ 
\int {\cal D}_{bc}\Psi^*_{bc}\Psi_{bc} \textsf{S}_b(\bb_b,x_0) 
\textsf{S}_c(\bb_c,x_0)\nonumber\\
&=&
\textsf{S}_a(\bb,x_0)+ \int {\cal D}_{bc}\Psi^*_{bc}\Psi_{bc} 
[ \textsf{S}_b(\bb_b,x_0) \textsf{S}_c(\bb_c,x_0)-\textsf{S}_a(\bb,x_0)],
\label{eq:3.B.7}
\eea
where in the last line we applied the completeness 
condition (\ref{eq:3.A.4}). It gives precisely 
the same result for $\delta \textsf{S}_a(\bb,x)$ as the unitarity condition. 
The technique (\ref{eq:3.B.7}) was applied earlier
to the derivation of the color dipole BFKL equation \cite{NZ94,NZZBFKL}.



\section{The calculation of the virtual radiative corrections to the leading
parton spectrum}


\subsection{Leading partons off the free-nucleon target}

With the allowance for the radiative correction, the quasielastic 
scattering matrix $1-\textsf{S}_{a}(\bb,x_0)$ in the master formula
(\ref{eq:2.B.8}) must be replaced by $1-\textsf{S}_{el}(\bb,x)$
and
\bea
[1-\textsf{S}_{el}(\bb,x)][1-\textsf{S}_{el}^\dagger(\bb',x)] &=&
[1-\textsf{S}_{a}(\bb,x_0)][1-\textsf{S}_{a}^\dagger(\bb',x_0)]
\nonumber\\
&-& \delta \textsf{S}_a(\bb,x)[1-\textsf{S}_{a}^\dagger(\bb',x_0)]
- [1-\textsf{S}_{a}(\bb,x_0)]\delta \textsf{S}_a^\dagger(\bb',x)
\label{eq:4.A.1}
\eea
Here first term leads to the Born spectrum. Upon using of the result
(\ref{eq:3.B.6}), the virtual correction, $d\sigma^{VRC}$, 
to the Born spectrum takes the form 
(hereafter we suppress the calculation of the free-nucleon and nuclear
matrix elements)
\bea
{d\sigma^{VRC}(x) \over d^2\bp} &&= -2\cdot  
\int {d^2\bb d^2\bc \over (2 \pi)^2} \exp(-i\bp\cdot\bc) 
\int {\cal D}_{bc}\Psi^*_{bc}\Psi_{bc} 
[ \textsf{S}_b(\bb_b,x_0) \textsf{S}_c(\bb_c,x_0) +
\nonumber \\
&&
\textsf{S}_a(\bb,x_0) \textsf{S}_a^\dagger(\bb-\bc,x_0)
-\textsf{S}_a(\bb,x_0) 
- \textsf{S}_b(\bb_b,x_0) 
\textsf{S}_c(\bb_c,x_0)\textsf{S}_a^\dagger(\bb,x_0)]
\nonumber\\
&&
= -
\int  {d^2\bc \over (2 \pi)^2} \exp(-i\bp\cdot\bc) 
 \int {\cal D}_{bc}\Psi^*_{bc}\Psi_{bc} 
[\sigma_{\bar{a}bc}(x_0) + \sigma_a(x_0) 
-\sigma_{bc}(x_0)-
\sigma_{a\bar{a}}(x_0)]
\label{eq:4.A.2}
\eea

The general technique of the derivation of the multiparton cross sections is found in 
\cite{NZ94,Nonlinear,QuarkGluonDijet,GluonGluonDijet}, here we only cite the results:
\bea
\sigma_{bc}(x) &=& {C_b + C_c - C_a \over 2C_F}\sigma(x,\brho)+ 
{C_a \over 2C_F}\sigma_0(x),
\nonumber\\
\sigma_{\bar{a}bc}(x) &=& {C_b + C_c - C_a \over 2C_F}
\sigma(x,\bb_b-\bb_c)
+{C_a + C_c - C_b \over 2C_F}\sigma(x,\bb'-\bb_c)
+ {C_a + C_b - C_c \over 2C_F}\sigma(x,\bb'-\bb_b)
\nonumber\\
&=&{C_b + C_c - C_a \over 2C_F}\sigma(x,\brho) 
+
{C_a + C_c - C_b \over 2C_F}\sigma(x,\bc +z_b\brho)
+
{C_a + C_b - C_c \over 2C_F}\sigma(x,\bc-z_c\brho).
\nonumber\\
\label{eq:4.A.3}
\eea
The integrand of (\ref{eq:4.A.2}) takes the form
\bea
&&\sigma_{\bar{a}bc}(x) + \sigma_a(x) -\sigma_{bc}(x)-
\sigma_{a\bar{a}}(x)=
\nonumber\\
&&={1\over 2C_F}[(C_a + C_c - C_b)\sigma(x,\bc +z_b\brho)
+ (C_a + C_b - C_c)\sigma(x,\bc-z_c\brho)
-2C_a\sigma(x,\bc)]
 \nonumber\\
&&={1\over 2C_F}\int d^2\bkappa f(x,\bkappa)\exp(i\bkappa\cdot\bc)
[2C_a - (C_a + C_c - C_b) \exp(iz_b\bkappa\cdot\brho)
- (C_a + C_b - C_c)\exp(-iz_c\bkappa\cdot\brho)].
\label{eq:4.A.4}
\nonumber \\
\eea

Now we specify the phase space limits. Let $x=\bp^2/W_{aN}^2$ and
the boundary condition be specified at $x_0$. Then the $z_c$ 
integration goes over $[z_{min},z_{max}]$, where 
$z_{max}=1-z_{min}$, and 
$ 
z_{max}z_{min} x_0 W_{aN}^2 = \bp^2$.
The resulting VRC to quasielastic scattering can be cast in two forms.
In the color dipole representation
\bea
{d\sigma^{VRC}(x,\bp)\over d^2\bp}&=& 
- {1\over 2C_F}\cdot {1\over (2\pi)^2}f(x_0,\bp) 
\int_{z_{min}}^{z_{max}} dz_c \int d^2\brho \Psi^*_{bc}(z_c,\brho)\Psi_{bc}(z_c,\brho) 
\nonumber\\
&\times&\Big[2C_a - (C_a + C_c - C_b)\exp(iz_b \bp \cdot\brho)-
(C_a + C_b - C_c)\exp(-iz_c\bp \cdot\brho)\Big]\nonumber\\
&=& -{d\sigma_{Qel}^B(x_0) \over d^2\bp}\cdot {C_a\over C_F(2\pi)^2}
\int_{x/x_0}^1 dz_c \int d^2\brho \Psi^*_{bc}(z_c,\brho)\Psi_{bc}(z_c,\brho) 
\nonumber\\
&\times&
\Big\{ [1 -\exp(iz_b\bp\cdot \brho)] +[1- \exp(-iz_c\bp\cdot\brho)]
+ ({C_A \over C_a}-1)[\exp(-iz_c\bp\cdot \brho)-\exp(iz_b\bp\cdot \brho)]
\Big\}
\nonumber \\
\label{eq:4.A.5}
\eea
has the expected form of the Born cross section times the combination of
form factors of the $bc$ Fock state. This combination of form factors
vanishes at $\bp=0$. In the final result we made an explicit use
of $C_b=C_a$ and $C_c=C_A$ which is the case in reactions
of the practical interest: $a\to ag$.

In the subsequent analysis of the small-$x$ evolution properties 
of the leading particle spectrum it would be more convenient to work
in the $k_\perp$-factorization representation
\bea
{d\sigma^{VRC}(x,\bp)\over d^2\bp}  
&=&f(x_0,\bp) {1\over 2C_F}\cdot{1\over 2(2\pi)^2}\int_{z_{min}}^{z_{max}}  dz_c 
\int d^2\bkappa
\Big[C_A |\Psi(z_c,\bkappa)-\Psi(z_c,\bkappa-z_b\bp)|^2 \nonumber\\
&+&
(2C_a  - C_A)|\Psi(z_c,\bkappa)-\Psi(z_c,\bkappa+z_c\bp)|^2\Big]\nonumber\\
&=&f(x_0,\bp) {1\over 2C_F}\cdot{1\over 2(2\pi)^2}\int_{x/x_0}^1 dz_c 
2\alpha_S P_{ca}(z_c)\nonumber\\
&\times&\int d^2\bkappa [C_A K(\bkappa,\bkappa-z_b\bp) +
(2C_a-C_A)K(\bkappa,\bkappa+z_c\bp)]
\label{eq:4.A.6}
\eea

\end{widetext}

The averaging over the initial, and summing
over the final, colors and polarizations of partons is encoded in  
$|\Psi(z_c,\bkappa)-\Psi(z_c,\bkappa-\bp)|^2$, for which
we use the explicit form \cite{SingleJet}
\bea
&&|\Psi(z_c,\bkappa)-\Psi(z_c,\bkappa-\bDelta)|^2 =
\nonumber \\
&=&
2\alpha_S P_{ca}(z_c)\left|{\bkappa \over \bkappa^2+\varepsilon^2}-
{\bkappa -\bDelta \over (\bkappa-\bDelta)^2+\varepsilon^2}\right|^2
\nonumber\\
&=& 2\alpha_S P_{ca}(z_c)K(\bkappa,\bkappa-\bDelta),
\label{eq:4.A.7}
\eea
where 
\beq
\varepsilon^2 = z_b z_c Q_a^2 + z_b m_c^2 + z_c m_b^2 \,,
\label{eq:4.A.8}
\eeq
$Q_a^2$ is the virtuality of the projectile parton defined by its 
transverse momentum in the incident proton, $Q_a^2=\bk_a^2$,
and $P_{ca}(z_c)$ is the real-emission part of the familiar 
splitting function. For incident quarks, $a=q$,
\bea
P_{ga}(z_g)&=&C_a {1+(1-z_g)^2 \over z_g},
\label{eq:4.A.9}
\eea
for incident gluons  one would use the splitting function 
$P_{gg}(z_g)$, which also will be relevant to the color octet incident quarks.
We emphasize in passing that in transitions $a\to bc$ the helicity of 
partons mixes with the orbital angular momentum, but that has no bearing on the
description of the polarization summed final states because the $s$-channel 
helicity of partons is conserved in the scattering
process.

In the calculation of the total radiative correction, the above VRC 
must be combined with 
the (nonlinear) $k_\perp$ factorization results for the 
real emission  \cite{SingleJet}. Making use of the
representation (\ref{eq:4.A.7}), the spectrum of 
color-triplet leading quarks, $b=a=q$, from the
real emission $a \to bg$ can be cast in the form ($\bp\equiv \bp_b$) 
\bea
&&\left.{d\sigma^{REC}(x,\bp)\over dz_b d^2\bp}\right|_{a \to bg}  = 
\nonumber \\
& =&{1\over 2C_F} \cdot{1\over 2 (2\pi)^2}
\cdot 2\alpha_S P_{ga}(z_g) \nonumber\\
&\times&
\int d^2\bkappa f(x_0,\bkappa)\Big\{ C_A [K(\bp,\bp-\bkappa) 
\nonumber \\
&+& K(\bp,\bp - z_b \bkappa)] 
\nonumber\\
&+& (2C_F -C_A)K(\bp-\bkappa,\bp - z_b \bkappa)\Big\}.
\label{eq:4.A.10}
\eea
The corresponding result for the color-octet incident partons is 
\bea 
&& \left.{d\sigma^{REC}(x,\bp )\over dz_b d^2\bp}\right|_{a \to bg} =
\nonumber \\ 
&=& {1\over 2C_F} {1 \over 2
(2\pi)^2} 2\alpha_S P_{ga}(z_g)\nonumber\\
&\times& C_A\int d^2\bkappa f(x_0,\bkappa)
 \Big\{
K(\bp,\bp-z_g\bkappa) 
\nonumber \\
& +&K(\bp-\bkappa,\bp-z_g\bkappa)
 +K(\bp,\bp-\bkappa) \Big\}.
\label{eq:4.A.11} 
\eea

Once the  finite-$z_g$ VRC and REC for quarks are known, we can proceed
to inclusive jets in collisions of protons. Here a crucial
point is a cancellation of 
effects of interactions of spectator partons of the beam proton 
in the inclusive single-particle cross section summed over
all excitations of the beam and target \cite{NPZcharm,SingleJet}.
Let $ q_a(x_N,\bk_a)$ be an unintegrated density of quarks of
flavor $a$,  carrying a fraction $x_a$ of the proton's 
lightcone momentum and having the transverse momentum $\bk_a$.
Then the inclusive spectrum of leading jets with the Feynman variable
$x_F$ and the transverse momentum $\bp$, produced by the 
interacting quark $a$   
 will be given by the standard convolution, cf. Eq. (\ref{eq:2.A.5}),
\begin{widetext}
\bea
&&{d\sigma_a(x_F,\bp )\over dx_F d^2\bp}=
\int_{x_F}^1 dx_a \int d^2\bk_a \int_0^1
 dz_b d^2\bp'\delta(x_F-z_b x_a)
\delta(\bp-\bp'-z_b\bk_a)\nonumber\\
&\times&  q_a(x_a,\bk_a) \Biggl\{\Biggl[
{d\sigma_{Qel}^B(x,\bp' )\over d^2\bp'}+
{d\sigma^{VRC}(x,\bp' )\over d^2\bp'}\Biggr]\delta(1-z_b)+
{d\sigma^{REC}(x,\bp' )\over dz_b d^2\bp'}\Biggl\},
\label{eq:4.A.12} 
\eea
where $
 x_a x W_{pN}^2= {\bp^2/z_b z_c}$ and 
 $W_{pN}$ is the proton-nucleon cms energy.



\subsection{Color-triplet  leading quarks off the nuclear target}

In the case of heavy nuclei one considers cross sections per
unit area in the impact parameter plane. Starting with 
Eq. (\ref{eq:4.A.2}), we calculate first the nuclear matrix 
element of the multiparton $\textsf{S}$-matrices. The resulting 
master formula for the VRC takes the form
\bea
{d\sigma^{VRC}(x,\bp) \over d^2\bp d^2\bb} &=& -2\cdot 
\int {d^2\bb d^2\bc \over (2 \pi)^2} \exp(-i\bp\cdot\bc) 
\int \Psi^*_{bc}\Psi_{bc} {\cal D}_{bc}
\Big\{\textsf{S}[\bb;\sigma_{bc}(x_0)] 
+\textsf{S}[\bb;\sigma_{a\bar{a}}(x_0)]
- \textsf{S}[\bb;\sigma_{a}(x_0)] 
\nonumber  \\
&& 
-\textsf{S}[\bb;\sigma_{\bar{a}bc}(x_0)] 
\Big\}
\label{eq:4.B.1}
\eea
For color-triplet 
incident quarks $C_a=C_b=C_F$. 
Making use of (\ref{eq:4.A.3}), and treating
the nucleus as the dilute uncorrelated gas of colorless nucleons,  
we readily find the product representation for multiparton $\textsf{S}$-matrices:
\bea
&&\textsf{S}[\bb;\sigma_{bc}(x_0)]
+\textsf{S}[\bb;\sigma_{a\bar{a}}(x_0)]
-
\textsf{S}[\bb;\sigma_{a}(x_0)]-\textsf{S}[\bb;\sigma_{\bar{a}bc}(x_0)]
\nonumber\\
&&=\Big\{\textsf{S}[\bb;{C_A\over 2C_F}\sigma(x_0,\brho)]
\textsf{S}[\bb;{1\over 2} \sigma_{0}(x_0)]-
\textsf{S}[\bb;{1\over 2} \sigma_{0}(x_0)]\Big\}\nonumber\\
&&+\Big\{\textsf{S}[\bb; \sigma(x_0,\bc)]
-
\textsf{S}[\bb;{C_A\over 2C_F}\sigma(x_0,\brho)]
\textsf{S}[\bb;{C_A\over 2C_F}\sigma(x_0,\bc+z_b\brho)]
\textsf{S}[\bb;{2C_F-C_A\over 2C_F}\sigma(x_0,\bc-z_g\brho)]\Big\}.
\nonumber \\
\label{eq:4.B.2}
\eea
The Fourier transform of the first group is of terms, which is 
$\bc$-independent,
will be $\propto \delta^{(2)}(\bp)$. We evaluate the second group 
to the leading order of the large-$N_c$
perturbation theory, when $C_A=2C_F$ and 
$\textsf{S}[\bb;{2C_F-C_A\over 2C_F}\sigma(x_0,\bc-z_g\brho)]=1$.
In this approximation it admits a simple
Fourier representation 
\bea
&&\textsf{S}[\bb; \sigma(x_0,\bc)]-
\textsf{S}[\bb;{C_A\over 2C_F}\sigma(x_0,\brho)]
\textsf{S}[\bb;{C_A\over 2C_F}\sigma(x_0,\bc+z_b\brho)]=
\nonumber\\
&=&\int d^2\bkappa \phi(\bb,x_0,\bkappa)\exp(i\bkappa\cdot \bc)
\times \Big\{1-\textsf{S}[\bb;\sigma(\brho)]\exp(iz_b \bkappa\cdot \brho)\Big\}
.
\label{eq:4.B.3}
\eea
The VRC splits naturally into two terms:
\bea
{d\sigma^{VRC}(x,\bp) \over d^2\bp d^2\bb} &=& 
2\delta^{(2)}(\bp)
\int_{z_{min}}^{z_{max}} dz_c d^2\brho \Psi^*_{bc}(z_c,\brho)\Psi_{bc}(z_c,\brho)\nonumber\\
&\times&\{1-\textsf{S}[\bb;\sigma(x_0,\brho)]\}\cdot
\textsf{S}[\bb;{1\over 2}\sigma_{0}(x_0)]\cdot \{1-
\textsf{S}[\bb;{1\over 2} \sigma_{0}(x_0)]\} 
\nonumber\\
&-&2\phi(\bb,x_0,\bp)
\int_{z_{min}}^{z_{max}} dz_c d^2\brho \Psi^*_{bc}(z_c,\brho)\Psi_{bc}(z_c,\brho)
\{1-\textsf{S}[\bb;\sigma(x_0,\brho)]
\exp(iz_b\bp\cdot\brho)\}.\nonumber\\
\label{eq:4.B.4}
\eea
\end{widetext}

Evidently, the term $\propto \delta^{(2)}(\bp)$ 
in the r.h.s. of (\ref{eq:4.B.4}) 
is the VRC to the elastic scattering. 
Indeed,  the VRC
$\delta\textsf{S}_{el}(\bb)$ to the elastic $\textsf{S}$-matrix
equals 
\bea
\delta \textsf{S}_{el}(\bb) &=& \bra{A} \delta \textsf{S}_a(\bb)\ket{A}
\nonumber \\
&=&
\int dz_c d^2\brho \Psi^*_{bc}(z_c,\brho)\Psi_{bc}(z_c,\brho)
\nonumber \\
&& \times 
\{\textsf{S}[\bb;\sigma_{bc}] -\textsf{S}[\bb;\sigma_a]\}.
\label{eq:4.B.5}
\eea
With the large-$N_c$ cross section (\ref{eq:4.A.3}), this gives 
\bea
\delta\textsf{S}_{el}(\bb) &=& -\textsf{S}[\bb;{1\over 2} \sigma_{0}(x_0)]
\nonumber\\
&\times& 
\int dz_c d^2\brho \Psi^*_{bc}(z_c,\brho)
\Psi_{bc}(z_c,\brho)
\nonumber \\
&\times& \{1-\textsf{S}[\bb;\sigma(x_0,\brho)]\}.
\label{eq:4.B.6}
\eea
Now, the VRC to the profile function of pure elastic scattering
can be evaluated from the expansion
\bea
&&\{ 1-\textsf{S}[\bb,{1\over 2}\sigma_{0}(x_0)]-
\delta\textsf{S}_{el}(\bb) \}^2\nonumber\\
&&=
\{ 1-\textsf{S}[\bb,{1\over 2}\sigma_{0}(x_0)]\}^2 
\nonumber \\
&-& 2\delta\textsf{S}_{el}(\bb) 
\{ 1-\textsf{S}[\bb,{1\over 2}\sigma_{0}(x_0)]\}.
\label{eq:4.B.7}
\eea
In conjunction with (\ref{eq:4.B.6}) one obtains precisely the 
integrand of the term $\propto \delta^{(2)}(\bp)$ in the r.h.s.
 of (\ref{eq:4.B.4}) 
which accomplishes the proof.

Now we isolate the nonlinear $k_\perp$ factorization
for the virtual correction to quasielastic scattering
of color-triplet quarks:
\bea
&& {(2\pi)^2d\sigma^{VRC}_{Qel}(x,\bp) \over d^2\bp d^2\bb} 
=\nonumber \\
&-& 2\phi(\bb,x_0,\bp)
\int_{z_{min}}^{z_{max}} dz_c d^2\brho \Psi^*_{bc}(z_c,\brho)\Psi_{bc}(z_c,\brho)\nonumber\\
&\times& \{1-\textsf{S}[\bb;\sigma(x_0,\brho)]
\exp(iz_b\bp\cdot\brho)\}
\nonumber\\ 
&=&-\phi(\bb,x_0,\bp)\int_{z_{min}}^{z_{max}} dz_c 2\alpha_S P_{gq}(z_c) \int d^2\bkappa_1
\nonumber\\
&\times &
\Big\{\textsf{S}[\bb;\sigma_{0}(x_0)]K(\bkappa_1,\bkappa_1-z_b\bp)
\nonumber\\
&+& 
\int d^2\bkappa_2 \phi(\bb,x_0,\bkappa_2)
K(\bkappa_1,\bkappa_1+\bkappa_2 -z_b\bp)\Big\}.
\nonumber \\
\label{eq:4.B.8}
\eea
Here we made use of the Fourier representation (\ref{eq:2.C.5}) for 
$\textsf{S}[\bb;\sigma(\brho)]$  and the sum rule (\ref{eq:2.C.6}).
In view of Eq.~(\ref{eq:2.C.11}), and in close similarity to the free-nucleon
case, the VRC has the expected form of 
the Born cross section times a $\bp$ dependent form factor.
The difference from the free-nucleon case is that this
from factor does not vanish at $\bp=0$.

The above VRC must be combined with REC derived in Ref. \cite{SingleJet} 
to the same leading order of the large-$N_c$ perturbation theory
($a=b=q$):
\bea 
&&\left.{ (2\pi)^2 d\sigma^{REC}(x,\bp) \over dz_b d^2\bp
d^2\bb}\right|_{q\to qg} = 2\alpha_S P_{gq}(z_g) \nonumber\\
&&\times \int d^2\bkappa_1 \phi(\bb,x_0,\bkappa_1)
\nonumber \\
&&
\Biggl\{\textsf{S}[\bb;\sigma_{0}(x_0)] 
[K(\bp,\bp-\bkappa)+ K(\bp,\bp-z_b \bkappa) ]\nonumber \\
&&+ \int d^2\bkappa_2 \phi(\bb,x_0,\bkappa_2)
K(\bp- \bkappa_1,\bp- z_b \bkappa_2)\Biggr\} . 
\nonumber \\
\label{eq:4.B.9}
\eea
For a practical calculation of the LPM effect and nuclear
quenching of leading quark jets off nuclei one would
use an obvious  generalization of the convolution (\ref{eq:4.A.12}) to
nuclear targets.


\subsection{Color-octet leading quarks and leading gluons off nuclei}

The case of color-octet incident partons gives still more insight into
the leading parton production. To the Born approximation, the spectrum
of color-octet leading partons (\ref{eq:2.C.11}) is described by the 
glue $\phi_{gg}(\bb,x_0,\bkappa)$ defined through the $\textsf{S}$-matrix for
the octet-octet color dipoles \cite{SingleJet}:
\bea
\Phi_{gg}(\bb,x_0,\bp) &&= \int {d^2\br \over (2 \pi)^2} \textsf{S}[\bb;{C_A\over C_F}
\sigma(x_0,\br)]  
\nonumber \\
&& \times \exp( -i\bp\cdot \br). 
\label{eq:4.C.1}
\eea
The issue is how this result changes if one includes the VRC and REC 
for gluons radiated  in the regime of nuclear coherency in the
rapidity span between the jet and the target nucleus.  

A tricky point is that in our derivation we shall encounter
not the collective glue (\ref{eq:4.C.1}) but still
another component of the color-space density matrix for their collective
glue.  Specifically, in the case of incident color-octet
partons  $C_a=C_b=C_c=C_A$
and all the multiparton dipole cross sections (\ref{eq:4.A.3}) 
are superpositions of
\bea
\sigma_g(x,\br)&=&{C_A\over 2C_F}\sigma(x,\br)={1\over 2}\sigma_{gg}(x,\br),\nonumber\\
\sigma_{g,0}(x)&=&{C_A\over 2C_F}\sigma_0(x) = {1\over 2}\sigma_{gg,0}(x).
\label{eq:4.C.2}
\eea
Correspondingly, the multiparton nuclear $\textsf{S}$-matrices will
be products of $\textsf{S}[\bb;\sigma_g(x,\bc)]$, 
and there emerges a collective nuclear glue $\Phi_g(\bb,x,\bkappa)$ such that
\beq
\textsf{S}[\bb;\sigma_g(x,\bc)]= 
\int d^2\bkappa \Phi_g(\bb,x,\bkappa) \exp(i\bkappa\cdot \bc).  
\label{eq:4.C.3}
\eeq
In contrast to $\Phi(\bb,x_0,\bp)$ and $\Phi_{gg}(\bb,x_0,\bp)$, the
so-defined auxiliary $\Phi_g(\bb,x_0,\bkappa)$ is a component of the color density
matrix for the collective nuclear glue which is not
directly measurable in the 
quasielastic scattering of partons.
   
\begin{widetext}
What would enter the master formula for color-octet leading partons
is 
\bea
&&\textsf{S}[\bb;\sigma_{bc}(x_0)]
+\textsf{S}[\bb;\sigma_{a\bar{a}}(x_0)]
-
\textsf{S}[\bb;\sigma_{a}(x_0)]-\textsf{S}[\bb;\sigma_{\bar{a}bc}(x_0)]\nonumber\\
&&
=\textsf{S}[\bb;\sigma_g(x_0,\brho)]\textsf{S}[\bb;\sigma_{g,0}(x_0)]
-
\textsf{S}[\bb;\sigma_{g,0}(x_0)]
+\textsf{S}[\bb; 2\sigma_g(x_0,\bc)]\nonumber \\
&&
- \textsf{S}[\bb;\sigma_g(x_0,\brho)]\textsf{S}[\bb;\sigma_g(x_0,\bc+z_b\brho)]
\textsf{S}[\bb;\sigma_g(x_0,\bc-z_g\brho)]\nonumber\\
&&=-\textsf{S}[\bb;\sigma_{g,0}(x_0)]\{1- \textsf{S}[\bb;\sigma_{g,0}]\} .
\{1- \textsf{S}[\bb;\sigma_{g}(x_0,\brho)]\}
\nonumber\\
&&
+
\textsf{S}[\bb;\sigma_{g,0}(x_0)]
\int d^2\bkappa \exp(i\bkappa \cdot \bc)\phi_g(\bb,x_0\bkappa)
\Bigl\{2 -\textsf{S}[\bb;\sigma_{g}(x_0,\brho)] \cdot 
[\exp(iz_b\bkappa \cdot \brho)+\exp(-iz_g \bkappa \cdot \brho)]\Bigr\}
\nonumber\\
&&
+\int d^2\bkappa_1 d^2\bkappa_2 \phi_g(\bb,x_0,\bkappa_1)\phi_g(\bb,x_0,\bkappa_2)
\exp(i(\bkappa_1+\bkappa_2) \cdot \bc) 
\nonumber \\
&&\times
\Big\{1- \textsf{S}[\bb;\sigma_{g}(x_0,\brho)]\
\exp[i\brho(z_b\bkappa_1-iz_g \bkappa_2)]\Big\}.
\label{eq:4.C.4}
\eea
Again we made an extensive use of the dilute gas treatment
of the nucleus.
Precisely as it was the case with the color-triplet quarks,
the $\bc$ independent term in (\ref{eq:4.C.4}) gives rise to the 
VRC to the pure elastic scattering:
\bea
{d\sigma^{VRC}_{el} \over d^2\bp d^2\bb} &=& 
\delta^{(2)}(\bp)\cdot 2\textsf{S}[\bb;\sigma_{g,0}(x_0)]\cdot 
\{1- \textsf{S}[\bb;\sigma_{g,0}(x_0)]\}
 \int dz_c d^2\brho \Psi^*_{bc}(z_c,\brho)\Psi_{bc}(z_c,\brho)
\{1-\textsf{S}[\bb;\sigma_g(x_0,\brho)]\}.\nonumber\\
\label{eq:4.C.5}
\eea
The  nonlinear $k_\perp$ factorization for the quasielastic 
scattering of the color-octet quarks is much more subtle:
 \bea
&&(2\pi)^2 {d\sigma^{VRC}(x,\bp) \over d^2\bp d^2\bb} = -
\int_{z_{min}}^{z_{max}} dz_g 2\alpha_S P_{ga}(z_g) \int d^2\bkappa_1
\nonumber\\
&\times& \Biggl\{
\phi_g(\bb,x_0,\bp)
\textsf{S}^2[\bb;\sigma_{g,0}(x_0)]
[K(\bkappa_1,\bkappa_1+ z_b\bp) + 
K(\bkappa_1,\bkappa_1+z_g\bp)]\nonumber\\
&-& \textsf{S}[\bb;\sigma_{g,0}(x_0)]\int d^2\bkappa_2
\Bigl\{\phi_g(\bb,x_0,\bp-\bkappa_2)\phi_g(\bb,x_0,\bkappa_2)
K(\bkappa_1,\bkappa_1-\bkappa_2+z_g\bp)\nonumber\\
&+& \phi_g(\bb,x_0,\bp)\phi_g(\bb,x_0,\bkappa_2)
[K(\bkappa_1,\bkappa_1+\bkappa_2+z_b\bp) 
+K(\bkappa_1,\bkappa_1+\bkappa_2+z_g\bp)
]\Big\}\nonumber\\
&-& \int d^2\bkappa_2 d^2\bkappa_3 \phi_g(\bb,x_0,\bp-\bkappa_2)
\phi_g(\bb,x_0,\bkappa_2)\phi_g(\bb,x_0,\bkappa_3)
K(\bkappa_1,\bkappa_1-\bkappa_2+\bkappa_3+z_g\bp)  \Biggr\}.
\label{eq:4.C.6}
\eea
In striking contrast to the case of color-triplet quarks, 
the VRC to quasielastic scattering of color-octet quarks is
no longer proportional to the Born approximation. 
In the calculation of the total spectrum of color-octet leading
quarks, the VRC (\ref{eq:4.C.6}) must be combined with the
REC derived in Ref. \cite{SingleJet}:
\bea 
&&\left.{(2\pi)^2 d\sigma^{REC}(x\bp) \over dz_b d^2\bp d^2\bb}
\right|_{a\to bg} =  2\alpha_S P_{ga}(z_g)\int d^2\bkappa_1 
\phi_g(\bb,x_0,\bkappa_1) \nonumber\\
&&\times\Biggl\{
\textsf{S}^2[\bb;\sigma_{g,0}(x_0)][
K(\bp,\bp+z_b\bkappa_1) + K(\bp +\bkappa,\bp+z_b\bkappa) + 
K(\bp,\bp+\bkappa) ] \nonumber \\
&&+ \textsf{S}[\bb;\sigma_{g,0}(x_0)] \int  d^2\bkappa_2 
\phi_g(\bb,x_0,\bkappa_2)\nonumber\\
&&\times [ K(\bp+\bkappa_1,\bp+z_b\bkappa_2)+K(\bp +\bkappa_1+\bkappa_2,
\bp+z_b\bkappa_2) + K\big(\bp+\bkappa_1,\bp+z_b(\bkappa_1+\bkappa_2)\big)]
\nonumber\\
&&+\int d^2\bkappa_2 d^2\bkappa_3
 \phi_g(\bb,x_0,\bkappa_2) \phi_g(\bb,x_0,\bkappa_3)K\big(\bp+\bkappa_1+\bkappa_3,
\bp+z_b(\bkappa_2+\bkappa_3)\big)  \Biggr\} . 
\label{eq:4.C.7} 
\eea
\end{widetext}


\subsection{The $s$-channel unitarity and the quark number sum rule}

Although our derivation of the virtual radiative corrections makes 
manifest use of the $s$-channel unitarity, it is useful to revisit 
this issue from the viewpoint of the leading quark-number sum rule.
The latter dictates that the leading quark inclusive cross section
must integrate to exactly the total quark-target interaction cross
section, i.e., the sum of the integrated virtual and real
radiative corrections to the single-quark inclusive cross section
must equal the increment of the total quark-target cross section 
caused by the gluon production in the $s$-channel. 

We start with the integrated real emission cross section. The integration
over the transverse momenta $\bp_c$ and $\bp_b$ in the master
formula (\ref{eq:3.A.8}) gives $\bb_c =\bb_c'$ and  $\bb_b =\bb_b'$,
which entails $\bb=\bb'$, $\brho=\brho'$. Then, the unitarity condition
(\ref{eq:3.A.12}) leads to $S^{(4)}_{\bar{b}\bar{c} c b} (\bb_b',\bb_c',\bb_b,\bb_c)=1$
and $\textsf{S}^{(2)}_{\bar{a}a}(\bb',\bb)=1$. The three-parton cross section
which enters $\textsf{S}_a^{\dagger}(\bb')\textsf{S}_b(\bb_b) \textsf{S}_c(\bb_c)$
would equal
\bea
\sigma_{\bar{a}bc}&=&{C_b + C_c - C_a \over 2C_F}\sigma(\brho)+
{C_a + C_c - C_b \over 2C_F}\sigma(z_b\brho) 
\nonumber \\
&&+
{C_a + C_b - C_c \over 2C_F}\sigma(z_c\brho).
\label{eq:4.D.1}
\eea
The result for the integrated real emission cross section is (we suppress
the matrix elements over the target nucleon and/or nucleus)
\bea
\delta\sigma^{REC} \Big( a^* \to b(\bp_b) c(\bp_c) \Big) &=&
2\int  {\cal D}_{bc} \Psi^*_{bc}\Psi_{bc} d^2\bb_c 
\nonumber \\
&& \times [1-\textsf{S}_{\bar{a}bc}], 
\label{eq:4.D.2}
\eea
cf. the derivation of open charm production in Ref. \cite{NPZcharm}.

The integration over the transverse momentum $\bp$ in the master formula 
(\ref{eq:4.A.2}) amounts to $\bc=0$ and
\bea
\delta\sigma^{VRC} &=& -2\int \Psi^*_{bc}\Psi_{bc} {\cal D}_{bc} d^2\bb_c 
\nonumber \\ &&\times [1+\textsf{S}_{bc}-\textsf{S}_{a}-\textsf{S}_{\bar{a}bc}]
\, .
\label{eq:4.D.3}
\eea

Then, the total radiative correction to the inclusive single-quark
cross section would equal
\bea
\delta\sigma^{REC+VRC}&=& 2\int \Psi^*_{bc}\Psi_{bc} {\cal D}_{bc} d^2\bb_c 
[\textsf{S}_{a}-\textsf{S}_{bc}]
\nonumber\\
&=& - 2\int \Psi^*_{bc}\Psi_{bc} {\cal D}_{bc} d^2\bb_c\delta
\textsf{S}_{a} 
\nonumber \\ 
&=& \delta\sigma_{tot} \, .
\label{eq:4.D.4}
\eea
In the last step we used our result (\ref{eq:4.B.3}). 
We emphasize that Eq. (\ref{eq:4.D.4})
gives the exact result for the increment of the
 total quark-gluon cross cross section
for radiation of one gluon, without resorting to the 
soft gluon, i.e., LL${1\over x  }$
approximation. The proof of the quark-number
sum rule is accomplished.


\subsection{A mini-summary on the LPM effect for leading jets and particle}


The nuclear modification of the radiation quenching 
(stopping) of leading
jets produced at large Feynman variable and fixed transverse
momentum -- the $\bp$-dependent LPM effect for leading jets --
is our principal new result. The nonlinear $k_\perp$ factorization 
of Secs. IV.B and IV.C gives the leading particle spectra in 
the form of explicit nonlinear quadratures in terms of the
collective nuclear glue.  
Such quadratures for both real and virtual pQCD radiative 
corrections are not found
in the previous works by other groups on forward jet production
\cite{Kovchegov,Blaizot,Fujii}. Still 
another novelty of our approach is a derivation
of the virtual pQCD radiative corrections from
the $s$-channel unitarity. Our approach to the solution of
the $s$-channel unitarity holds for both free nucleon and
nuclear targets. The emerging interplay of the real and
virtual radiative corrections to the leading jet spectra
is precisely the same as in the evolution equation for
the total cross section which property
guarantees a manifest fulfillment of the 
leading parton number sum rule. 
For color octet (adjoint representation) 
leading quarks and gluons
our  nonlinear $k_\perp$ factorization holds for
arbitrary $N_c$, only for color-triplet (fundamental
representation) leading quarks one would invoke
large-$N_c$ perturbation theory for the formulation of
closed-form quadratures. 


\section{Linear $k_\perp$ factorization for the leading parton spectra
to LL${1\over x}$?}


\subsection{The master formula for LL${1\over x}$ evolution of the color-dipole $\textsf{S}$-matrix and the 
collective nuclear glue beyond the Born approximation}


For further insight into the properties of 
leading parton spectra, in this section we consider the 
LPM effect-deconvoluted 
cross sections, i.e., the spectra evaluated in the LL${1\over x}$,
i.e., soft gluon, approximation
neglecting the radiative energy loss. The 
relatively simple case of the free-nucleon
target will be considered in the next subsection,
only with the nuclear targets one encounters nontrivial conceptual issues. 
Here, motivated by the linear
$k_\perp$-factorization (\ref{eq:2.C.11}) observed to the Born approximation,
we would like to compare the  LL${1\over x}$ evolution properties of the 
leading particle spectra and of the collective nuclear glue. We recall 
that in Ref. \cite{NSSdijet,NSSdijetJETPLett,Nonlinear} we defined the
collective nuclear glue in terms of the amplitude for coherent
diffractive dijets. At arbitrarily small $x$ this observable is 
calculable through the color-dipole $ \textsf{S}$-matrix 
$\textsf{S}_{a\bar{a}}(x,\bb_a,\bb_{\bar{a}})$. At  a boundary $x_0=x_A$, i.e., 
at the onset of coherent
nuclear effects, $\textsf{S}_{a\bar{a}}(x_0,\bb_a,\bb_{\bar{a}})$ is given 
by Eq. (\ref{eq:2.C.2}) which defines the Born approximation for the collective
glue (\ref{eq:2.C.5}). We would like to maintain the definition 
(\ref{eq:2.C.5}) for arbitrary values of $x$ and need the   LL${1\over x}$ 
evolution properties of the color dipole  $\textsf{S}$-matrix.
For soft gluons, $z_g \ll 1$, there are several important simplifications.
First, the impact parameters
of partons which radiate soft gluons, do not change, see Eq.~(\ref{eq:3.A.5}).
The resulting master formula for the small-$x$ renormalization of the
color-dipole  $\textsf{S}$-matrix for the perturbative $a\bar{a}g$  
Fock state in the $a\bar{a}$-dipole is an
obvious generalization of Eq. (\ref{eq:3.B.6}) \cite{NZ94,NZZBFKL}:
\begin{widetext}
\beq
\delta \textsf{S}_{a\bar{a}}(x,\bb_a,\bb_{\bar{a}}) = 
\int_{x/ x_0}^1  dz_g \int d^2\brho \Psi^*_{a\bar{a}g}\Psi_{a\bar{a}g}
[ \textsf{S}_a(x_0,\bb_a)  \textsf{S}_a^\dagger(x_0,\bb_{\bar{a}})
\textsf{S}_g(x_0,\bb_g)-
 \textsf{S}_{a\bar{a}}(x_0,\bb_a,\bb_{\bar{a}})].
\label{eq:5.A.1}
\eeq
\end{widetext}
Second, in view of Eq. (\ref{eq:4.A.8}), the parameter $\varepsilon^2$ 
in the $ag$ wave function would not 
depend on the virtuality of partons $a,\bar{a}$ in the $a\bar{a}$ color
dipole. Consequently, the distribution of gluons in the $a\bar{a}$
dipole will be
given by \cite{NZ94,NZZBFKL}  
\beq
\Psi^*_{a\bar{a}g}\Psi_{a\bar{a}g} = |\Psi_{ag}(z_g,\brho)- \Psi_{ag}(z_g,\brho+\br)|^2,
\label{eq:5.A.2}
\eeq
where  $\br=\bb_a-\bb_{\bar{a}}$, $\brho=\bb_g-\bb_a$. In the momentum space
\beq
\Psi_{ag}(z_g,\bk) \propto {\bk \over \bk^2 +\mu_g^2}, 
\label{eq:5.A.3} 
\eeq
where we indicated the (optional) infrared regularization parameter $
\mu_g$ which models the finite 
propagation radius of perturbative gluons \cite{NZZBFKL,Shuryak}. The 
corresponding small-$z_g$ splitting function equals
\beq
P_{ga}(z_g)= {2C_a \over z_g}. 
\label{eq:5.A.4}
\eeq
To the  LL${1\over x}$ we have $[z_{min},z_{max}]=[x/x_0,1]$ and the  
$z_g$ integration would give the familiar LL${1\over x}$
factor 
\beq
\int_{x/x_0}^1 {dz_g\over z_g}= \log{x_0\over x}. 
\label{eq:5.A.5}
\eeq

\begin{widetext}

\subsection{Free-nucleon target: BFKL evolution for the leading parton spectrum}

We start with the free-nucleon target. We first recall the connection between
the LL${1\over x}$ evolution equation for color dipole cross 
section \cite{NZ94,NZZBFKL} and
the BFKL equation \cite{KLF,BFKL} for the unintegrated glue of the nucleon. 

In terms of the dipole cross sections, Eq. (\ref{eq:5.A.1}) amounts to the 
evolution correction \cite{NZ94,NZZBFKL}


\bea
\delta \sigma_{a\bar{a}}(x,\br) &=& \int_{x/x_0}^1 dz_g \int d^2\brho
 |\Psi_{ag}(z_g,\brho)- \Psi_{ag}(z_g,\brho+\br)|^2 
\big(\sigma_{a\bar{a}g}(x_0)-\sigma_{a\bar{a}}(x_0) \big)
\nonumber\\
&=& {C_A \over 2 C_F}\int dz_g \int d^2\brho
 |\Psi_{ag}(z_g,\brho)- \Psi_{ag}(z_g,\brho+\br)|^2 \nonumber\\
&\times& 
\big[ \sigma(x_0,\brho) +\sigma(x_0,\brho+\br) -\sigma(x_0,\br) \big],
\label{eq:5.B.1}
\eea
and to the evolution of the unintegrated glue $\delta f(x,\bp)$
\beq
\int {d^2\br \over (2 \pi)^2} \delta\sigma_{a\bar{a}}(x,\br)\exp(-i\bp\cdot \br)
= \delta\sigma_{a\bar{a},0}(x)\delta^{(2)}(\bp)-{C_a\over C_F}\delta f(x,\bp).
\label{eq:5.B.2}
\eeq
In the transition from the color-dipole to the momentum representation 
we follow the familiar route. 
First, we make use of
\bea 
&&\sigma(x,\brho) +\sigma(x,\brho+\br) -\sigma(x,\br)=
\int d^2\bkappa f(x,\bkappa)[1-\exp(i\bkappa\cdot\brho)][1+\exp(i\bkappa\cdot\br)].
\label{eq:5.B.3}
\eea
Second, the symmetry with respect to $\brho \leftrightarrow \brho+\br$ allows 
a simplifying substitution of $|\Psi_{ag}(z_g,\brho)- \Psi_{ag}(z_g,\brho+\br)|^2$ by
$2\Psi_{ag}^*(z_g,\brho)[\Psi_{ag}(z_g,\brho)- \Psi_{ag}(z_g,\brho+\br)]$ which
has a Fourier representation
\bea
\Psi_{ag}^*(z_g,\brho)[\Psi_{ag}(z_g,\brho)-
\Psi_{ag}(z_g,\brho+\br)] &=&
\int d^2\bkappa_1 d^2\bkappa_2 \Psi^*(z_g,\bkappa_2)\Psi(z_g,\bkappa_1)
\exp[i(\bkappa_1-\bkappa_2)\cdot\brho]
\nonumber \\ &&
\times [1-\exp(i\bkappa_1\cdot\br)].
\label{eq:5.B.4}
\eea
Then the Fourier transform (\ref{eq:5.B.2}) gives 
\bea
{C_a\over C_F}\delta f(x,\bp) &=& {C_A\over C_F}\cdot {1\over (2\pi)^2}
\int_{x\over x_0}^1  dz_g \int d^2\bkappa d^2\bkappa_1 d^2\bkappa_2 
f(x_0,\bkappa) \Psi^*(z_g,\bkappa_2)\Psi(z_g,\bkappa_1)
\nonumber\\
&\times&
[\delta^{(2)}(\bkappa_1-\bkappa_2) -\delta^{(2)}(\bkappa+\bkappa_1-\bkappa_2)]\nonumber\\
&\times&
[\delta^{(2)}(\bkappa_1-\bp)+\delta^{(2)}(\bkappa+\bkappa_1-\bp)-
\delta^{(2)}(\bkappa-\bp)]\nonumber\\
&=& 
{C_A\over C_F}\cdot {1\over (2\pi)^2}
\int_x^1  dz_g  2\alpha_S P_{ga}(z_g)\nonumber\\
&\times& 
 \int d^2\bkappa \big\{2K(\bp,\bp-\bkappa) f(x_0,\bkappa)-
f(x_0,\bp)K(\bkappa,\bkappa-\bp) \big\}
\label{eq:5.B.5}
\eea
Upon the use of the soft-gluon approximation (\ref{eq:5.A.4}) , the
differential form of Eq. (\ref{eq:5.B.5}) boils down to
precisely the BFKL equation for the unintegrated gluon density 
of the target nucleon \cite{KLF,BFKL}:
\bea
{\partial f(x,\bp)\over
\partial \log{1\over x}} &=& 
{\cal K}_0 \int d^2\bkappa 
\big[2K(\bp,\bp-\bkappa)f(x,\bkappa)-f(x,\bp)K(\bkappa,\bkappa-\bp) \big] =
\Bigl({\cal K}_{BFKL}\otimes f\Bigr)(x,\bp)
\label{eq:5.B.6}
\eea
\end{widetext}
where 
\beq
{\cal K}_0={C_A\over 2\pi^2}\alpha_S.
\label{eq:5.B.7}
\eeq
The manifest dependence on the Casimir $C_a$ of projectile
partons in the lhs of Eq. (\ref{eq:5.B.5}) dipole cancels against $C_a$ in the
$a\to ag$ splitting function. Similarly, we obtain
\bea
{\partial \delta\sigma_{a\bar{a},0}(x)\over
\partial \log{1\over x}}
&=&{C_a\over C_F}{\cal K}_0\int d^2\bq d^2\bkappa  
K(\bq,\bq+\bkappa)f(x,\bkappa)
\nonumber \\
&=& {C_a\over C_F}\int d^2\bkappa {\partial f(x,\bkappa)\over
\partial \log{1\over x}}.\label{eq:5.B.8}
\eea
Note the consistency between equations (\ref{eq:5.B.8}) and 
(\ref{eq:5.B.6}).

Now we proceed to the LL${1\over x}$ approximation 
for quasielastic scattering. 
In the REC the energy loss by the quasielastically 
scattered parton is neglected,
i.e., one must put  $z_b=1$ where legitimate 
and integrate over $z_g$ with the soft-gluon 
approximation (\ref{eq:5.A.4}). The contribution to the integrand of
(\ref{eq:4.A.10}) from
the term $\propto (2C_F -C_A)K(\bp-\bkappa,\bp - z_b \bkappa)$
vanishes. Similarly, the contribution to the VRC from the term $\propto 
(2C_a - C_A) K(\bkappa,\bkappa+z_c\bp)$ vanishes at $z_c=z_g \ll 1$. 
Combining together the virtual and real corrections and the Born cross
section, we find the $z$-integrated inclusive $\bp$ spectrum of leading
partons 

\bea
&&{d\sigma_{Qel}(x,\bp) \over d^2\bp}=
{C_a\over 2C_F}\Biggl\{
f(x_0,\bp)+ 
{\cal K}_0 \log{x_0\over x} \int d^2\bkappa \nonumber\\
&\times&
[2K(\bp,\bp-\bkappa)f(x_0,\bkappa)-
f(x_0,\bp)K(\bkappa,\bkappa-\bp)]\Biggr\}\nonumber\\
&=& {C_a\over 2C_F}\Biggl\{
f(x_0,\bp)+ 
\log{x_0\over x} \Bigl({\cal K}_{BFKL}\otimes f \Bigr)(x_0,\bp)\Biggr\}
\nonumber \\
&=& {C_a\over 2C_F}f(x,\bp)
\nonumber\\
&=&{d\sigma_{Qel}(x_0,\bp) \over d^2\bp}+ 
\log{x_0\over x} \Bigl({\cal K}_{BFKL}\otimes {d\sigma_{Qel} 
\over d^2\bp}\Bigr)(x_0,\bp).\nonumber\\
\label{eq:5.B.9}
\eea

We conclude, that to the LL${1\over x}$ approximation, the inclusive spectrum of 
leading partons maps the BFKL-evolved 
unintegrated glue in the target nucleon, as it was announced in the Introduction,
Eq. (\ref{eq:1.1}). This spectrum by itself satisfies the linear BFKL evolution!
It is proportional to
the Casimir of the projectile parton $C_a$, i.e., satisfies the
expected Regge factorization. Of course, these Regge factorization 
properties are precisely what
one would have expected from the dominance of multiregge production 
processes to LL${1\over x}$ \cite{KLF,BFKL}. In our approach, the VRC
component of the BFKL kernel -- the origin of the reggeization of 
gluons -- derives from the manifest imposition of 
the $s$-channel unitarity; such a derivation is of obvious
relevance to the case of nuclear targets with severe unitarity constraints.

\subsection{The first LL${1\over x}$ iteration for 
nuclear glue: triplet-antitriplet $q\bar{q}$ dipoles and 
color-triplet leading quarks}

The principal reasoning behind our definition \cite{NSSdijet,NSSdijetJETPLett}
of the collective 
nuclear glue -- the linear $k_\perp$-factorization for the amplitude 
of excitation of coherent diffractive dijets -- remains valid beyond
the nuclear Born approximation at $x\approx x_A$,
see below Sec. V.E. Consequently, we stick to (\ref{eq:2.C.5})
as the all-$x$ 
definition  of the $x$-dependent collective nuclear
unintegrated glue $\Phi(\bb,x,\bp)$. 
We need to specify the LL${1\over x}$ evolution properties of this
$\textsf{S}$-matrix. As we saw already at the Born approximation,
the collective nuclear glue is a density matrix in the color space.
We need to investigate how the  LL${1\over x}$ evolution depends
on the color multiplet the partons of the color-dipole belong to.

We start with the first iteration of the LL${1\over x}$
evolution for the triplet-antitriplet color dipole $\textsf{S}$-matrix 
to the leading order of the large-$N_c$ perturbation theory
in which $C_A=2C_F$.
The effect of the perturbative $q\bar{q}g$ Fock state in the
$q\bar{q}$ color dipole is given by a nuclear
generalization of Eq.~(\ref{eq:5.A.1}):
\begin{widetext}
\bea
\delta \textsf{S}_{q\bar{q}}(\bb;x,\br) &=& \int dz_g \int d^2\brho
 |\Psi_{ag}(z_g,\brho)- \Psi_{ag}(z_g,\brho+\br)|^2 
\Big\{\textsf{S}[\bb;\sigma_{q\bar{q}g}(x_0)]-
\textsf{S}[\bb;\sigma_{0}(x_0)]\Big\} \nonumber\\
&=& \int dz_g \int d^2\brho
 |\Psi_{ag}(z_g,\brho)- \Psi_{ag}(z_g,\brho+\br)|^2 \nonumber\\
&\times& \Big\{ \textsf{S}[\bb;\sigma(x_0,\brho)] 
\textsf{S}[\bb;\sigma(x_0,\brho+\br)]-
\textsf{S}[\bb;\sigma(x_0,\br)]\Big\},
\label{eq:5.C.1}
\eea
\end{widetext}
where we made use of the large-$N_c$ approximation for $\sigma_{q\bar{q}g}$
of Eq. (\ref{eq:4.A.3}) (for the early suggestions to consider (\ref{eq:5.C.1})
as the closed-form equation for the color dipole $\textsf{S}$-matrix see
Refs. \cite{BalitskyNonlinear,KovchegovNonlinear}, the recent
dispute and further references are found in Refs. \cite{Trianta,Armesto}). 
The first LL${1\over x}$ iteration for the collective nuclear glue 
$\delta\phi(\bb,x,\bp)$ is defined by 
\bea
\delta\Phi(\bb,x,\bp)&=& 
\int {d^2\br \over (2 \pi)^2} \delta \textsf{S}_{q\bar{q}}(\bb,x,\br)
\exp(-i\bp\cdot \br) \nonumber\\
&=& \delta S_{q\bar{q},0}(\bb,x)\delta^{(2)}(\bp) +
\delta\phi(\bb,x,\bp)
\nonumber \\
\label{eq:5.C.2}
\eea
Following the technique exposed in Sec. IV, one readily
finds the nonlinear $k_\perp$-factorization quadrature 
(the preliminary results were reported
at conferences \cite{NonlinearBFKL})
\bea
&&\delta\Phi(\bb,x,\bp)=\int_{x/x_0}^1 dz_g 2\alpha_S 
P_{gq}(z_g)\int {d^2\bkappa_1 d^2\bkappa_2 \over (2 \pi)^2}
\nonumber\\
&\times&\Biggl\{
 \Phi(\bb,x_0,\bkappa_1)\Phi(\bb,x_0,\bkappa_2)
K(\bp+\bkappa_1,\bp+\bkappa_2)\nonumber\\
&-&\int d^2\bkappa_1 d^2\bkappa_2 
\Phi(\bb,x_0,\bp)\Phi(\bb,x_0,\bkappa_2)
\nonumber \\
&&
\times K(\bkappa_1,\bkappa_1+\bkappa_2) \Biggr\}.
\label{eq:5.C.3}
\eea
Upon the application of the expansion (\ref{eq:2.C.5}), this equation
splits into
\bea
&& \delta S_{q\bar{q},0}(\bb,x) 
=  - \textsf{S}[\bb;\sigma_0(x_0)]{\cal K}_0 \log{x_0\over x} 
\nonumber\\
 && \times \int d^2\bkappa d^2\bq 
K(\bq, \bq + \bkappa )\phi(\bb,x_0,\bkappa),
\label{eq:5.C.4}
\eea
where we notice a nice correspondence to the result (\ref{eq:5.B.8})
for the free-nucleon target, and 
\bea
\delta\phi(\bb,x,\bp)& = & {\cal K}_0 \log{x_0\over x}
\Biggl\{\textsf{S}[\bb;\sigma_0(x_0)]\int  d^2\bkappa
\nonumber\\
&\times& \Big[2K(\bp,\bp-\bkappa)\phi(\bb,x_0,\bkappa) -
\nonumber \\
&&\phi(\bb,x_0,\bp)K(\bkappa,\bkappa-\bp)\Big]
\nonumber\\
&+&\int d^2\bkappa_1 d^2\bkappa_2 
\phi(\bb,x_0,\bkappa_2)
\nonumber\\
&\times&
\Big[ K(\bp+\bkappa_1,\bp+\bkappa_2)\phi(\bb,x_0,\bkappa_1) 
\nonumber \\ 
&&
-
 K(\bkappa_1,\bp+\bkappa_1+\bkappa_2)\phi(\bb,x_0,\bp)\Big]\Biggr\}.
\nonumber \\
\label{eq:5.C.5}
\eea
The evolved collective nuclear glue equals
\beq
\phi(\bb,x,\bp)= \phi(\bb,x_0,\bp)+\delta\phi(\bb,x,\bp).
\label{eq:5.C.6}
\eeq
Notice, that the result (\ref{eq:5.C.5})  
contains the linear component which evolves with the conventional
BFKL kernel, but this component is suppressed by the nuclear
attenuation factor $\textsf{S}[\bb;\sigma_{0}(x_0)]$ (see, however,
below Sec. V.G). If we identify the collective glue $\phi(\bb,x_0,\bp)$
with the exchange by a nuclear pomeron, then the absorption-non-suppressed
component of (\ref{eq:5.C.5}) can be viewed as a fusion of two nuclear
pomerons.

Now we proceed to the first LL${1\over x}$ correction to the quasielastic 
scattering of color-triplet quarks. As in the case of the
free-nucleon target, it must be evaluated for $z_c=z_g\ll 1$ and $z_b=1$
with soft-gluon approximation (\ref{eq:5.A.4}) for the quark
splitting function.
Combining together the virtual and real corrections we find 
(the negative valued terms originate from VRC)
\begin{widetext}
\bea
{d\delta\sigma_{Qel}(x,\bp) \over  d^2\bb d^2\bp} &=&  {\cal K}_0 \log{x_0\over x}
\Biggl\{\textsf{S}[\bb;\sigma_{0}(x_0)]\int  d^2\bkappa
\Big[2K(\bp,\bp-\bkappa)\phi(\bb,x_0,\bkappa)-
\phi(\bb,x_0,\bp)K(\bkappa,\bkappa-\bp)\Big]\nonumber\\
&+&\int d^2\bkappa_1 d^2\bkappa_2 \phi(\bb,x_0,\bkappa_2)
\nonumber\\
&\times&
\Big[ K(\bp+\bkappa_1,\bp+\bkappa_2)\phi(\bb,x_0,\bkappa_1)-
 K(\bkappa_1,\bp+\bkappa_1+\bkappa_2)\phi(\bb,x_0,\bp)\Big]\Biggr\}\nonumber\\
&=&\delta\phi(\bb,x,\bp).
\label{eq:5.C.7}
\eea
\end{widetext}
In conjunction with the Born contribution (\ref{eq:2.C.11}), we obtain the 
LL${1\over x}$ evolved quasielastic scattering spectrum 
\bea
{d\sigma_{Qel}(x,\bp) \over  d^2\bb d^2\bp} = \phi(\bb,x,\bp),
\label{eq:5.C.8}
\eea
which is 
linear factorizable in terms of the 
collective nuclear unintegrated glue as defined through the 
excitation of coherent diffractive dijets.
By hindsight, we attribute
this finding to the Abelianization of the single-jet problem to the 
LL${1\over x}$ approximation, for the related
discussion see Ref. \cite{SingleJet}.


\subsection{The first LL${1\over x}$ iteration for 
nuclear glue: octet-octet color dipoles and color-octet leading 
quarks (gluons)}

Still more insight into the properties of inclusive production of 
leading partons comes from the inclusive spectra of leading
color-octet partons off nuclear targets. In this case there is
no need to invoke the large-$N_c$ approximation. One 
defines the octet-octet (gluon-gluon) collective nuclear
glue $\Phi_{gg}(\bb,x,\bp)$ as the Fourier transform
of the nuclear $\textsf{S}$-matrix
for octet-octet color dipoles. 
The Born approximation, $\Phi_{gg}(\bb,x_0,\bp)$, is given 
by Eq.~(\ref{eq:4.C.1}). We also
recall the convolution property \cite{SingleJet}
\beq
\Phi_{gg}(\bb,x_0,\bp) =\Biggl(\Phi_{g}\otimes \Phi_{g}\Biggr)(x_0,\bp)
\label{eq:5.D.1}
\eeq
to hold to the Born approximation.

\begin{widetext}
The first LL${1\over x}$
iteration for the octet-octet  nuclear $\textsf{S}$-matrix 
reads 
\bea
\delta \textsf{S}_{gg}(\bb;x,\br) &=& \int dz_g \int d^2\brho
 |\Psi_{ag}(z_g,\brho)- \Psi_{ag}(z_g,\brho+\br)|^2  
\Big\{\textsf{S}[\bb;\sigma_{ggg}(x_0)]-
\textsf{S}[\bb;\sigma_{gg}(x_0)]\Big\} 
\nonumber\\
&=& \int dz_g \int d^2\brho
 |\Psi_{ag}(z_g,\brho)- \Psi_{ag}(z_g,\brho+\br)|^2 
\nonumber\\
&\times& 
\Big\{ \textsf{S}[\bb;\sigma_g(x_0,\brho)] 
\textsf{S}[\bb;\sigma_g(x_0,\brho+\br)]
\textsf{S}[\bb;\sigma_g(x_0,\br)]
-
\textsf{S}[\bb;\sigma_{gg}(x_0,\br)]\Big\}.
\label{eq:5.D.2}
\eea
Proceeding the now familiar route, one would find
\bea
&&\delta \Phi_{gg}(\bb,x,\bp)=2{\cal K}_0 \log{x_0\over x}
\nonumber\\
&&
\times \Biggl\{
\int d^2\bkappa_3 d^2\bkappa_1 d^2\bkappa_2 
\Phi_g(\bb,x_0,\bp-\bkappa_3)\Phi_g(\bb,x_0,\bkappa_1)\Phi_g(\bb,x_0,\bkappa_2)
K(\bkappa_3+\bkappa_1,\bkappa_3+\bkappa_2)\nonumber\\
&&
-\int d^2\bkappa_3 d^2\bkappa_1 d^2\bkappa_2 
\Phi_g(\bb,x_0,\bp-\bkappa_3)\Phi_g(\bb,x_0,\bkappa_3)\Phi_g(\bb,x_0,\bkappa_2)
K(\bkappa_1,\bkappa_3+\bkappa_1+\bkappa_2)\Biggr\}\nonumber\\
&&
= {\cal K}_0 \log{x_0\over x}\Biggl\{ 
-\textsf{S}^2[\bb;\sigma_{g,0}(x_0)]\delta^{(2)}(\bp) \int d^2\bkappa_1 d^2\bkappa_2 
\phi_g(\bb,x_0,\bkappa_1)K(\bkappa_2,\bkappa_1+\bkappa_2)\nonumber\\
&&+\textsf{S}^2[\bb;\sigma_{g,0}(x_0)]
\int d^2\bkappa \Big[2K(\bp,\bp+\bkappa) \phi_g(\bb,x_0,\bkappa)
-\phi_g(\bb,x_0,\bp)K(\bkappa,\bp+\bkappa)\Big]
\nonumber\\
&&+\textsf{S}[\bb;\sigma_{g,0}(x_0)]\int d^2\bkappa_1 d^2\bkappa_2
\Big[\phi_g(\bb,x_0,\bkappa_1)
\phi_g(\bb,x_0,\bkappa_2)K(\bp+\bkappa_1,\bp+\bkappa_2)
\nonumber\\
&&+ 2\phi_g(\bb,x_0,\bkappa_1)
\phi_g(\bb,x_0,\bkappa_2)K(\bp+\bkappa_1,\bp+\bkappa_1+\bkappa_2)
-
 \phi_g(\bb,x_0,\bp)         
\phi_g(\bb,x_0,\bkappa_2)K(\bkappa_1,\bkappa_1+\bkappa_2)
\nonumber\\
&&- 
 \phi_g(\bb,x_0,\bp-\bkappa_2)
\phi_g(\bb,x_0,\bkappa_2)K(\bkappa_1,\bkappa_1+\bkappa_2)\Big]
\nonumber\\
&&+
\int d^2\bkappa_3 d^2\bkappa_1 d^2\bkappa_2 \Big[
\phi_g(\bb,x_0,\bp-\bkappa_3)\phi_g(\bb,x_0,\bkappa_1)\phi_g(\bb,x_0,\bkappa_2)
K(\bkappa_3+\bkappa_1,\bkappa_3+\bkappa_2)\nonumber\\
&&-\int d^2\bq d^2\bkappa_1 d^2\bkappa_2 \phi_g(\bb,x_0,\bp-\bkappa_3)
\phi_g(\bb,x_0,\bkappa_3)\phi_g(\bb,x_0,\bkappa_2)
K(\bkappa_1,\bkappa_3+\bkappa_1+\bkappa_2)\Big]\Biggr\}\nonumber\\
&&=\delta\textsf{S}_{gg,0}(\bb,x)\delta^{2}(\bp) +\delta\phi_{gg}(\bb,x,\bp).
\label{eq:5.D.3}
\eea
\end{widetext}

The LL${1\over x}$ evolution of the the triplet-antitriplet collective 
nuclear glue $\Phi(\bb,x,\bp), \phi(\bb,x,\bp)$ is a (quadratic)
nonlinear 
$k_\perp$-factorization quadrature (\ref{eq:5.C.3}), (\ref{eq:5.C.5})
in terms of the same quantity. In striking contrast to that, 
for octet-octet color dipoles
$\delta \Phi_{gg}(\bb,x,\bp)$ is a (cubic) nonlinear 
$k_\perp$-factorization quadrature of another quantity --
$\Phi_{g}(\bb,x,\bp)$. The nuclear absorption-non-suppressed component of
(\ref{eq:5.D.3}) can be viewed as a fusion of three nuclear pomerons
described by $\Phi_{g}(\bb,x,\bp)$ to a different nuclear
pomeron  described by $\Phi_{gg}(\bb,x,\bp)$

The above distinction between the pomeron fusion properties of the 
nonlinear BFKL evolution for the triplet-antitriplet and octet-octet 
dipoles was not discussed before.
Now we notice an important unifying aspect of the 
LL${1\over x}$ evolution properties of  $\Phi(\bb,x,\bp)$ and 
$ \Phi_{gg}(\bb,x,\bp)$. Namely, the same answer for 
$\delta \Phi_{gg}(\bb,x,\bp)$  could 
have been derived from the already available results for the
triplet-antitriplet color dipole. 
Indeed, notice the factorization property
\begin{widetext}
\bea
&&\textsf{S}[\bb;\sigma_g(x_0,\brho)] \textsf{S}[\bb;\sigma_g(x_0,\brho+\br)]
\textsf{S}[\bb;\sigma_g(x_0,\br)]-
\textsf{S}[\bb;\sigma_{gg}(x_0,\br)]=
\nonumber\\
&&=\textsf{S}[\bb;\sigma_g(x_0,\br)]
\Big\{ \textsf{S}[\bb;\sigma_g(x_0,\brho)] \textsf{S}[\bb;\sigma_g(x_0,\brho+\br)]-
\textsf{S}[\bb;\sigma_g(x_0,\br)]\Big\}.
\label{eq:5.D.4}
\eea
\end{widetext}
The quantity in curly braces in the last line of (\ref{eq:5.D.4}) is
identical to the integrand of (\ref{eq:5.C.1}) subject to a simple
substitution 
\beq
\textsf{S}[\bb;\sigma(\bc)]\Rightarrow \textsf{S}[\bb;\sigma_g(\bc)].
\label{eq:5.D.5}
\eeq
Let $\delta\Phi_g(\bb,x,\bp)$ be the result of the iteration (\ref{eq:5.C.1}) subject
to the substitution (\ref{eq:5.D.5}). 
Then the factorization property (\ref{eq:5.D.4})
amounts to the convolution representation
\bea
\delta\Phi_{gg}(\bb,x,\bp)& = & 
2\Bigl(\Phi_{g}(\bb,x_0) \otimes \delta \Phi_{g}(\bb,x)\Bigr)(\bp).
\nonumber \\
\label{eq:5.D.6}
\eea 
The factor 2 in the rhs of Eqs. (\ref{eq:5.D.3}), (\ref{eq:5.D.6})
comes from the ratio 
of Casimirs in the $q\to qg$ and $g\to gg$ splitting functions.
Then (\ref{eq:5.D.6}) suggests that the evolved $\Phi_{gg}(\bb,x,\bp)$ is a
self-convolution (\ref{eq:5.D.1}) of the evolved $\Phi_{g}(\bb,x,\bp)$:
\bea
\Phi_{gg}(\bb,x,\bp)&=&\Phi_{gg}(\bb,x,\bp)+ \delta\Phi_{gg}(\bb,x,\bp)
\nonumber\\
&=&\Biggl([\Phi_{g}(\bb,x_0)+\delta\Phi_{g}(\bb,x)]
\nonumber \\
&&\otimes 
[\Phi_{g}(\bb,x_0)+\delta\Phi_{g}(\bb,x)]\Biggr)(\bp)
\nonumber\\
&=&
\Biggl(\Phi_{g}\otimes \Phi_{g}\Biggr)(\bb,x,\bp).
\label{eq:5.D.7}
\eea 
Recall that to the leading order of large-$N_c$ perturbation theory
$\Phi_{g}(\bb,x,\bp)=\Phi(\bb,x,\bp)$. Then Eqs. (\ref{eq:5.C.3}), 
(\ref{eq:5.C.5}) in conjunction with the convolution (\ref{eq:5.D.7})
provide a unified description of the LL${1\over x}$ evolution for
both the triplet-triplet and octet-octet collective nuclear glue.

A short comment on $\delta \textsf{S}_{gg,0}(\bb)$ is in order. Because
$\textsf{S}^2[\bb;\sigma_{g,0}(x_0)]=\textsf{S}[\bb;\sigma_{gg,0}(x_0)]$, it can
be cast in the form
\bea
\delta {\textsf{S}}_{gg,0}(\bb)=-
2{\textsf{S}}[\bb;\sigma_{gg,0}(x_0)]{\cal K}_0 \log{x_0\over x} 
\nonumber \\
\times \int d^2\bkappa_1 d^2\bkappa_2 
\phi_g(\bb,x_0,\bkappa_1)K(\bkappa_2,\bkappa_1+\bkappa_2).
\nonumber \\
\label{eq:5.D.8}
\eea 
Apart from the factor 2, the origin of which has been
explained following Eq. ~(\ref{eq:5.D.6}), it is an exact counterpart of 
Eqs.~(\ref{eq:5.B.8}) and (\ref{eq:5.C.4}).

We leave it as an exercise to check that when put together, the LL${1\over x}$
approximations for VRC, Eq. (\ref{eq:4.C.6}), and REC, Eq. (\ref{eq:4.C.7}),
give precisely
\bea
{d\sigma_{Qel} \over d^2\bb d^2\bp} = \phi_{gg}(\bb,x,\bp).
\label{eq:5.D.9}
\eea
We again recover the linear $k_\perp$-factorization for the 
quasielastic scattering spectrum in terms of a judiciously chosen
collective nuclear glue. Basically, this property is made obvious by
Eqs. (\ref{eq:5.D.7}) and (\ref{eq:5.C.8}).


\subsection{Going beyond the first iteration of the LL${1\over x}$ evolution?}

Will the exciting finding of linear $k_\perp$-factorization for 
quasielastic scattering of partons off nuclei hold to all orders
of the LL${1\over x}$ evolution? Will the leading jets offer
a long sought linear mapping of the collective nuclear glue? 

As far
as further iterations of Eqs. (\ref{eq:5.C.1}) and (\ref{eq:5.D.2})
are concerned the answer is a negative one - such a BK approximation 
can not be treated as a closed form equation. Still here we present 
heuristic arguments which partially vindicate the BK approximation
-- our scrutiny of all steps behind the derivation of  
Eqs. (\ref{eq:5.C.1}) and (\ref{eq:5.D.2}) suggests, that 
a limited number of
iterations of these equations will be a good working approximation
capable of describing evolution effects in the limited range of
energies attainable at RHIC and LHC.
Our line of reasoning is as follows. 

We consider the scattering of a
thin projectile (hadron) off a thick nucleus. The thickness of
the target nucleus is measured by a large parameter
\beq
\nu_A(\bb,x_0)={1\over 2}\sigma_0(x_0)T(\bb),
\label{eq:5.E.1}
\eeq
which is a number of nucleons in the tube of nuclear matter
of cross section ${1\over 2}\sigma_0(x_0)$ at an impact
parameter $\bb$. For instance, realistic models for
the color dipole cross section suggest $\sigma_0(x_0)\sim 60$ mb
\cite{BFKLregge} and   for central collisions with a gold
nucleus one finds $\nu_{Au}(\bb=0,x_0)\sim 6$.
We treat nuclear effects to all orders in
this large parameter $\nu_A(\bb,x_0)$. The coherent nuclear
effects start at $x_0=x_A$. At this boundary the nucleus can
be treated as a dilute gas of color-singlet nucleons
and  the resummation of nuclear effects boils down to
the Glauber-Gribov exponentials for multiparton 
$\textsf{S}$-matrices.

Recall that inelastic interactions with the free-nucleon 
target are described to
the single-gluon exchange in the $t$-channel.
The LL${1\over x}$ evolution is treated
as an effect of $ag$ Fock states in the projectile parton $a$
or $a\bar{a}g$ Fock states in the color dipole $a\bar{a}$.
Here emerges sort of a renormalization group (RG): upon the integration
over gluon variables, one would cast the interaction of the three-parton 
Fock state $a\bar{a}g$ in the form of an interaction of
the $a\bar{a}$ dipole. The sole effect of 
integrating out the gluon is a
renormalization of the color-dipole cross section, which is
equivalent to a renormalization of the gluon-density of the 
target. I.e., at each step of the  LL${1\over x}$ evolution,
the gluon from the $a\bar{a}g$ Fock state of the beam is 
RG reshuffled to become a part of the target,
while the beam is always described by the lowest Fock
state $a\bar{a}$. The result is the closed-form linear BFKL 
evolution in either the color dipole \cite{NZ94,NZZBFKL} 
or the unintegrated glue \cite{KLF,BFKL} representations reviewed
in Sec. V.B. Of course, to the single-gluon $t$-channel
exchange for inelastic processes, this renormalization procedure is
a manifestly beam-target symmetric one \cite{NZZBFKLJETP,NZZspectrum}.

In our discussion of the LL${1\over x}$ evolution for
nuclear targets we followed exactly the 
same RG arguments:  we reabsorbed the effect 
of the $a\bar{a}g$ Fock state of the beam into the
renormalization of the 
nuclear color-dipole $\textsf{S}$-matrix and the 
corresponding 
evolution of the collective glue $\Phi_{a\bar{a}}(\bb,x,\bp)$
of the target nucleus. The   color-dipole 
$\textsf{S}$-matrix and its Fourier transform - the collective glue -
are well defined to all orders in LL${1\over x}$
and are perfectly experimentally accessible: the 
frequently mentioned amplitude of excitation
of coherent hard diffractive dijets $\pi A \to q(\bp)\bar{q}(-\bp) A$ 
equals
\cite{NSSdijet,NSSdijetJETPLett,NZsplit} (we cite it for forward excitation)
\bea
{\cal M}(x,z_q,\bp) &=& 
\int {d^2\bb \over (2 \pi)^2} \int d^2\br \Psi_{\pi}(z_q,\br) 
\nonumber\\
&& \times  \exp(-i\bp \br)
[1-S_{q\bar{q}}(\bb,x,\br)] \nonumber \\ 
&=& \int d^2\bb \int d^2\bkappa \Big[\Psi_{\pi}(z_q,\bp) 
\nonumber\\
&& -
\Psi_{\pi}(z_q,\bkappa)\Phi_{q\bar{q}}(\bb,x,\bp-\bkappa) \Big] 
\nonumber\\
&\approx&  \Biggl\{\int d^2\bkappa \Psi_{\pi}(z_q,\bkappa)
\Biggr\} 
\nonumber \\ 
&&\cdot 
 \int d^2\bb \phi_{q\bar{q}}(\bb,x,\bp),
\label{eq:5.E.2}
\eea
where the last approximation holds for jets with the 
transverse momentum much larger than the intrinsic momentum
of quarks in the pion. 
Such a direct measurement of  $\phi_{q\bar{q}}(\bb,x,\bp)$ is possible
at arbitrarily small values of $x=M_{q\bar{q}}^2/W_{pN}^2$; 
a useful review
of the current experimental situation is found in \cite{Ashery}. 
In principle, the octet-octet collective glue $\phi_{gg}(\bb,x,\bp)$
is equally measurable in the diffractive breakup of glueballs. 

Then, what makes Eqs. (\ref{eq:5.C.1}) and (\ref{eq:5.D.2}) suspect
from the RG point of view? What we do is 
the adding of 
one more radiated gluon to the final state and an evaluation of
the matching radiative correction to the lower order cross section.
In the operator form, the master formula (\ref{eq:3.A.8}),
the identification (\ref{eq:3.A.9}) of the multiparton $\textsf{S}$-matrices,
the whole discussion of the unitarity driven renormalization 
(\ref{eq:3.B.6}), (\ref{eq:5.C.1})
and (\ref{eq:5.D.2}) of elastic $\textsf{S}$-matrices hold at an
arbitrary $x\lsim x_A$. The only caveat is the Glauber-Gribov 
factorization for nuclear matrix elements of the multiparton
$\textsf{S}$-matrices. Specifically, in all our derivations we
made an indiscriminate use of the property known to all practitioners
in the Glauber-Gribov multiple scattering theory \cite{Glauber,Gribov},
\beq
\bra{A} \prod_{i=1}^n\textsf{S}(\bb,\sigma_i)\ket{A}=
 \prod_{i=1}^n \bra{A}\textsf{S}(\bb,\sigma_i)\ket{A}.
\label{eq:5.E.3}
\eeq
It derives from (i) the nuclear $\textsf{S}$-matrix being
a product of the free-nucleon ones,
\beq
\textsf{S}_A(\bb)= \prod_{j=1}^A \textsf{S}_N(\bb-\bb_j),
\label{eq:5.E.4}
\eeq
where $\bb_j$ is the impact parameter of bound nucleon $j$,
and (ii) the dilute uncorrelated gas approximation for the 
nucleus.

Each RG reshuffling of the $s$-channel
gluon of the $a\bar{a}g$ Fock state from the beam to a
parton of the target nucleus introduces in the tube of
$\nu_A(\bb)$ nucleons a gluon which interacts with all
nucleons of the tube and breaks the assumption of the 
uncorrelated gas. Specifically, the Glauber-Gribov
factorization
\bea
\textsf{S}[\bb,\sigma_{q\bar{q}g}(x_0,\brho,\br)]&=&
 \textsf{S}[\bb;\sigma(x_0,\brho)] \nonumber\\
&\times&
\textsf{S}[\bb;\sigma(x_0,\brho+\br)],\nonumber\\
\textsf{S}[\bb,\sigma_{ggg}(x_0,\brho,\br)]&=&
 \textsf{S}[\bb;\sigma_g(x_0,\br)\textsf{S}[\bb;\sigma_g(x_0,\brho)] \nonumber\\
&\times&
\textsf{S}[\bb;\sigma_g(x_0,\brho+\br)],\nonumber\\
\label{eq:5.E.5}
\eea
holds only at the boundary $x_0=x_A$, i.e., for one $s$-channel
gluon, $n_g=1$. It will break down at smaller
$x$, beyond the first iteration, i.e., for $n_g\geq 2$.
While it prevents one from treating the BK approximation
as a closed form equation for the free-nucleon target
(for the related recent discussion see Ref. 
\cite{Trianta,Armesto}
and references therein), arguably, the 
breaking effects are small if the
multiplicity of RG gluons is small compared to 
the number $\nu_A(\bb)$ nucleons of in the tube of 
cross section $\half \sigma_0(x)$.
To be more precise, the number $n_g$ of iterations of the
BK approximation must be bounded from above: 
\beq
n_g  -1 \ll \nu_A(\bb).  
\label{eq:5.E.6}
\eeq

One can quantify $n_g$ from the experimentally observed small-$x$ 
growth of the unintegrated 
glue in the nucleons and/or the DIS structure function. In the fully 
developed BFKL asymptotics
\beq
F_{2p}(x,Q^2) \propto \left({1\over x}\right)^{\Delta_{\Pom}}=
\exp(\Delta_{\Pom}Y),
\label{eq:5.E.7}
\eeq
where $Y=\log{1\over x}$ and $\Delta_{\Pom}$ is the intercept of
the QCD pomeron. The expansion
\beq
\exp(\Delta_{\Pom}Y) =\sum_{n_g} {1\over n_g !}(\Delta_{\Pom}Y)^{n_g}
\label{eq:5.E.8}
\eeq
is an expansion over the multiplicity $n_g$ of $s$-channel gluons. The
average multiplicity equals
$\langle n_g \rangle = \Delta_{\Pom}Y$ and the average rapidity 
gap between gluons equals $\Delta Y = 1/\Delta_{\Pom}$.
Increasing the total rapidity $Y$ by $\Delta Y$ amounts to 
increasing $\langle n_g \rangle$ by unity and, simultaneously, 
to the increase of the structure function by the factor $e^1$. 
The experimental 
data on $\Delta_{\Pom}$ from HERA exhibit a steady
rise of $\Delta_{\Pom}$ from $\sim 0.1 $ at small $Q^2$ 
to $\sim  0.2$ at $Q^2\sim 10$GeV $^2$ to $\sim 0.3$
at $Q^2 \sim 100$ GeV $^2$ \cite{HERAdeltaPom}. 
For heavy nuclei $x_A\sim 10^{-2}$, which corresponds to
the nucleus fragmentation rapidity span $\eta_A\sim \log{1\over x_A} \sim
4.5$. 
If $Y$ is the total rapidity span,
then 
\beq
n_g-1 \approx \Delta_{\Pom}(Y-\eta_A)-1.
\label{eq:5.E.9}
\eeq
At LHC, for jets with $p_\perp \sim 10$ GeV the total kinematical 
rapidity span $Y \sim 15$, which entails an estimate  $n_g-1\lsim 2$.
For 
mini-jets with $p_\perp \sim 3$ GeV Eq. (\ref{eq:5.E.9}) gives
still smaller $n_g-1\lsim 1.5$. Comparing that to the above estimate for
$\nu_{Au}$, we conclude that several first 
iterations of the BK approximation would reproduce the gross
features of nuclear effects in $pA$ collisions in the energy range from
RHIC to LHC. 

The practical application to radiative stopping of 
( or the 
transverse momentum dependent LPM effect for )
leading
jets in $pA$ collisions at LHC would be as follows. Order
the $s$-channel gluons $g_1...g_{n}$
in energy in the nucleus rest 
frame: $z_1 < z_2 ...< z_{n}$. The stopping will be
dominated by the hard gluon $g_{n}$, the remaining
$n-1$  gluons 
carrying small energy would have a negligible impact on the 
energy loss and their principal effect would be the 
LL${1\over x}$ evolution of the collective
nuclear glue. Then the so-evolved collective nuclear glue
must be used in the formalism of Sec. IV as an input for 
the quantitative evaluation of the stopping of
leading quark jets caused by radiation of 
the hard gluon $g_{n}$.

Our discussion was limited to a scattering of a dilute and 
small projectile on a dense and large target.  Still it has 
a broad range of applicability 
at energies of RHIC and mid-rapidity production at LHC.
Indeed,  in this energy range proton
proton interactions which are still far away from the 
strong absorption regime of $\sigma_{el}/\sigma_{tot} = 1/2$
and a single-pomeron exchange for scattering of a dilute projectile on a 
dilute target is a
reasonable starting point \cite{BlockCahn,KaidalovAbs}.
 An extension to multipomeron exchanges
in proton-proton interactions and inclusion of pomeron
loops remain a major challenge in the field 
(for a review see \cite{Trianta}).    


\subsection{Extension to the LL${1\over x}$ treatment of
mid-rapidity production}

While our principal motivation has been the
interplay of real and virtual radiative corrections in the
LPM effect and quenching of leading high-$p_\perp$ 
jets, the extension of the above  LL${1\over x}$ 
considerations to mid-rapidity high-$p_\perp$ jets in DIS or 
in proton-nucleus collisions at LHC is straightforward. Here 
to LL${1\over x}$ the 
familiar linear $k_\perp$-factorization would hold on the beam 
side -- the discussion of cancellations of spectator 
interaction effects in the inclusive jet cross section
integrated over the phase space of spectator partons of the
beam is found in Refs. \cite{NPZcharm,SingleJet}. We also recall 
that to the Born approximation, the nonlinear  
$k_\perp$-factorization simplifies, for  LL${1\over x}$
real soft gluon
emission processes, to the linear one \cite{SingleJet,KovchegovMueller}. 
To the next order in pQCD, one must consider
the effect of real radiation in the rapidity span $[\eta_A,\eta]$
between the   nucleus fragmentation region and the
mid-rapidity jet, and include the corresponding virtual correction
to the Born cross section. 

We recall first the Born spectrum at the boundary $x_0=x_A$
for mid-rapidity gluons with transverse momentum
$\bp$, (pseudo)rapidity $\eta_g = \log(z/x)$
and $z=\bp^2/xW_{aN}\ll 1$ 
from  pQCD subprocesses $q\to qg$ and $g\to gg$ \cite{SingleJet}
\bea
&&{ (2\pi)^2 d\sigma_A\over d\eta_g d^2\bp
d^2\bb}\Biggr|_{a\to ag} = \nonumber\\
&&z\int d^2\bkappa \phi_{gg}(\bb,x_A,\bkappa)|\Psi(z,\bp) -
\Psi(z,\bp-\bkappa)|^2.\nonumber\\
\label{eq:5.F.1}
\eea
For the transverse momenta above the infrared parameter 
$\mu_g$, in (\ref{eq:4.A.7}) we can use
\beq
K(\bp,\bp-\bkappa)={\bkappa^2 \over \bp^2(\bp-\bkappa)^2},
\label{eq:5.F.2}
\eeq
identify the unintegrated glue in the 
incident parton $a$,
\beq
{dG_a (z,\bp-\bkappa) \over d^2\bp} = 
2\alpha_S zP_{ga}(z)\cdot{1\over (\bp-\bkappa)^2},
\label{eq:5.F.3}
\eeq
make use of (\ref{eq:2.C.11})
and cast (\ref{eq:5.F.1}) in the form
\bea
&&{ (2\pi)^2 d\sigma_A\over d\eta_g d^2\bp
d^2\bb}\Biggr|_{a\to ag} =\int d^2\bkappa d^2\bp_a 
\delta(\bp -\bp_a-\bkappa) \nonumber\\
&&\times
{dG_a (z,\bp_a) \over  d^2\bp_a}\cdot {\bkappa^2 \over \bp^2}\cdot
{d\sigma_{Qel}(x_A,\bkappa) \over d^2\bkappa}\Big|_{g\to g'}\nonumber\\
&&=  {4\pi \alpha_s\over N_c \bp^2}\int d^2\bkappa d^2\bp_a 
\delta(\bp -\bp_a-\bkappa) \nonumber\\
&&\times
{dG_a (z,\bp_a) \over  d^2\bp_a}\cdot {dG_{A,gg}(\bb,x_A,\bkappa)\over d^2\bb d^2\bkappa}
.
\label{eq:5.F.4}
\eea
The reabsorption of $\bkappa^2$ into
the definition 
\bea
\bkappa^2\phi_{gg}(\bb,x_A,\bkappa)= {4\pi \alpha_s\over N_c}\cdot  
{dG_{A,gg}(\bb,x_A,\bkappa)\over d^2\bb d^2\bkappa},\nonumber\\
\label{eq:5.F.5}
\eea 
brings (\ref{eq:5.F.4}) to a lucid  beam-target symmetric form.
The remaining factor $1/\bp^2$ is familiar from the square of 
the BFKL 
gluon radiation vertex \cite{KLF,BFKL} -- an absence of 
nuclear renormalization of this vertex is noteworthy. 

Modulo to the above explained factor $\bkappa^2/\bp^2$, 
Eq.~(\ref{eq:5.F.4}) is an exact 
counterpart of the convolution (\ref{eq:4.A.12}).
Upon this identification the differential cross section of 
quasielastic gluon-nucleon scattering, the further analysis 
of radiative corrections  would not
be any different from that presented above for leading jets. 
An important simplification is that the major effect of 
real radiation is a slight shift of the rapidity of 
the high-$p_\perp$ jet towards the nucleus region. In view
of the approximate boost invariance, this rapidity shift can 
safely be neglected. Then one is allowed to evaluate the real
and virtual radiative corrections in the LL${1\over x}$ 
approximation. Next we reiterate the above point that although
the real radiative corrections are highly nonlinear
quadratures of the collective nuclear glue \cite{SingleJet},
upon lumping together the real and virtual corrections all
nonlinear effects combine to the nonlinear LL${1\over x}$ evolution 
of the collective nuclear glue: $\phi_{gg}(\bb,x_A,\bkappa)$
of the Born approximation changes to the 
nonlinear LL${1\over x}$-evolved $\phi_{gg}(\bb,x,\bkappa)$.

\begin{widetext}

\mbox{}

\begin{figure}[!h]
\begin{center}
\includegraphics[width = 8cm,angle=270]{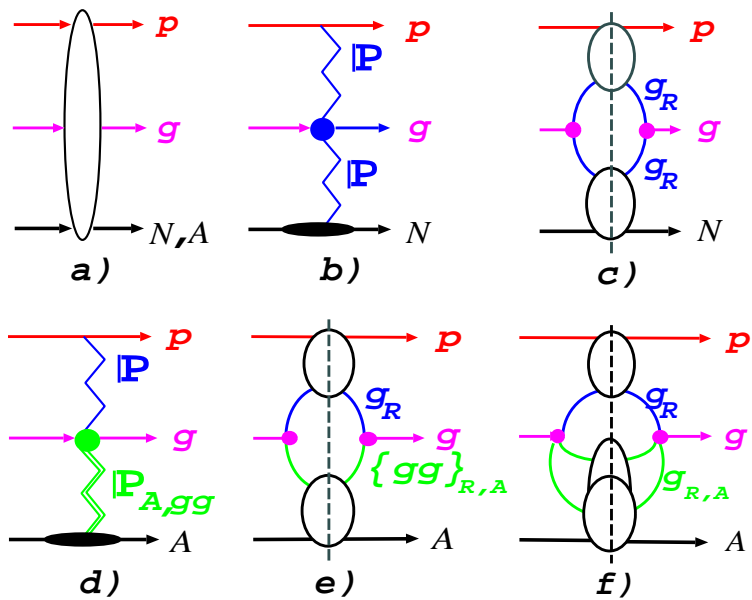}
\caption{\label{fig:KMglue2}
(a) The Kancheli-Mueller diagram for the single-gluon spectrum, 
(b) its pomeron exchange representation for mid-rapidity gluon
spectrum
in proton-nucleon collisions,
(c) its explicit form in terms of the emission of real gluons
off reggeized gluons $g_R$ and free-nucleon gluon density representation for 
the cut pomeron, (d) the pomeron exchange representation for 
mid-rapidity gluon spectrum in proton-nucleus collisions,
(c) the explicit form of the diagram (d) in terms of the emission 
of real gluons with transition between reggeized gluon $g_R$
and effective nuclear reggeized gluon $\{gg\}_{R,A}$ and free-nucleon 
and collective nuclear glue,(f) the representation of the
diagram (d) in terms of two cut nuclear pomerons $\Pom_{A,g}$
composed of effective nuclear reggeized gluons $g_{R,A}$.}
\end{center}
\end{figure}
\mbox{}
\end{widetext}

The beam-target symmetric
linear $k_\perp$ factorization form of Eq. (\ref{eq:5.F.4}) shows 
that sort of a LL${1\over x}$ 
BFKL multiregge factorization \cite{KLF,BFKL}
holds for the mid-rapidity single-jet spectrum off nuclei.
From the viewpoint of the Kancheli-Mueller $3\to 3$ 
optical theorem  \cite{KancheliMueller,SingleJet} of 
Fig. \ref{fig:KMglue2}, this form of the single-jet spectrum
suggests an interesting effective reggeon field theory 
interpretation (whether this interpretation and
the following conjectures will lead to a viable 
diagram technique \cite{LipatovAction} or not
needs further scrutiny). In the single mid-rapidity jet 
production on nucleons  the free-nucleon BFKL cut pomerons 
$\Pom$ are exchanged between the beam (target) and 
produced gluon (Fig.  \ref{fig:KMglue2}b), the more detailed
structure of this diagram in terms of the radiation of the 
gluon from reggeized gluons $g_R$ is shown in Fig. \ref{fig:KMglue2}c.
For the nuclear target, Fig.  \ref{fig:KMglue2}d, 
on the nucleus side one exchanges the cut
nuclear pomeron $\Pom_{A,gg}$ 
described by ${dG_{A,gg}(\bb,x,\bkappa)/ d^2\bb d^2\bkappa}$
which resums all multipomeron exchanges enhanced
by a large thickness of the nucleus. 
If we associate with this collective  glue the
effective nuclear reggeized gluon $\{gg\}_{R,A}$, then this 
Kancheli-Mueller diagram can be cast in the same form,
Fig.  \ref{fig:KMglue2}e, as in the free nucleon case,
Fig.  \ref{fig:KMglue2}c.
Our notation $\{gg\}_{R,A}$ for this effective gluon is a reminder
that it can be viewed as a composite state of two gluons $g_{R,A}$
associated with still another member of the family 
of nuclear pomerons, $\Pom_{A,g}$,
described by the collective glue  
${dG_{A,g}(\bb,x,\bkappa)/ d^2\bb d^2\bkappa}$ related
by Eq. (\ref{eq:5.F.5}) to $\Phi_g(\bb,x,\bkappa)$.
In due turn, the latter can be viewed as sort of 
a coherent composite state of the free-nucleon reggeized gluons.
Considering the distinct, nonlinear vs. linear,  
LL${1\over x}$ evolution properties of the free-nucleon 
and collective nuclear glue, 
the possibility of such a resummation and representation
in terms of a single nuclear reggeized  gluon $\{gg\}_{R,A}$ 
is far from being an obvious one. ( From the color dipole 
perspective it has its origin in the cancellation 
of spectator interactions and Abelianization of intranuclear
evolution of dipoles in the single jet problem.)

 The no-nuclear-renormalization of
the BFKL gluon emission vertex is a still another
nontrivial property: the $g_R g_R g$ and
$g_R \{gg\}_{R,A}g$ vertices are identical. Furthermore,
since  $\{gg\}_{R,A}$ is a composite state $\ket{g_{R,A} g_{R,A}}$,
the nuclear Kancheli-Mueller diagram of Fig. \ref{fig:KMglue2}d 
is a triple-pomeron diagram $\Pom \to \Pom_{A,g}\Pom_{A,g}$ with
three cut pomerons ( Fig. \ref{fig:KMglue2}f), and the
corresponding fully cut triple-pomeron vertex is not
renormalized by nuclear effects and is related to the 
square of the BFKL vertex. If we recall that $g_{R,A}$ is by itself  a
coherent composite state of reggeized gluons $g_R$,
and invoke our experience with unitarity rules for
dijet and single-jet production \cite{NSZZ_Gribov}, then 
the above non-renormalization property would apparently extend 
to a whole family of fully cut multipomeron  vertices
$\Pom \to (n \Pom) (m\Pom)$ in the single-gluon production
problem. A very different derivation of non-renormalization of
the gluon emission vertex in proton-nucleus 
interaction, based on the 
QCD shock wave scattering approach, is found in 
\cite{BalitskyShockWave}, although Balitsky 
considered only the Born approximation and did not
discuss the relevant interplay of the real and
virtual pQCD radiative corrections.

As we cited above, this multiregge, and linear
$k_\perp$-factorizable,  form of the mid-rapidity
gluon spectrum has already made an appearance
in the literature (\cite{SingleJet,KovchegovMueller,BalitskyShockWave}
and references therein).
A useful aspect of our analysis is an elucidation of
the r\^ole of the $s$-channel unitarity in the 
evaluation of the virtual pQCD radiative corrections
for nuclear targets. Our technique gives a closed form
nonlinear $k_\perp$ factorization quadratures for the spectra
without invoking reggeon field theory and
Abramovsky-Gribov-Kancheli (AGK) unitarity cutting rules
\cite{AGK}.
From the jet phenomenology
viewpoint the exchange by a nuclear pomeron on the nucleus
side of the Kancheli-Mueller diagram entails  
a full fledged familiar Cronin effect: nuclear attenuation for 
the transverse momenta below the nuclear saturation 
scale, $\bp^2 \lsim Q_A^2$, antishadowing for 
$\bp^2 \sim (2-3) Q_A^2$ and slow approach to the 
impulse approximation at large $\bp^2$, controlled by 
a nuclear higher twist correction $\propto  Q_A^2/\bp^2$.
A full discussion of the hot disputed issue of AGK
unitarity 
connections between diffractive and inelastic processes 
goes beyond the scope of the present communication
(the subtleties of multiple cut pomeron exchanges in pQCD
were commented on in \cite{NSZZ_Gribov} and will be
reported elsewhere \cite{NSZ_AGK}).
Here we only mention that one version of the AGK rules 
discussed in \cite{CapellaAGK} suggests the substitution
$\phi_{gg}(\bb,x,\bkappa)\to {1\over 2}T(\bb)f(x,\bkappa)$.
This amounts to the impulse approximation and misses the 
Cronin effect: numerically the departure from the impulse 
approximation is quite substantial (see Ref. \cite{SingleJet} 
and references therein). While our interest is in heavy targets, 
in his  AGK discussion of single gluon spectra 
Braun focused on the transition from 2 to 4 
$t$-channel gluons relevant to the triple-pomeron
diagram for the double scattering in the deuteron
(\cite{Braun3P,Braun2006}) and references therein).
As far as the nuclear
departure from the impulse approximation is concerned 
his results for double scattering are similar to ours
for multiple scattering for heavy nuclei. In the reggeon 
field theory formalism used by Braun 
the LL${1\over x}$ virtual pQCD corrections are reabsorbed into the
Regge trajectory of the gluon. We still find 
our manifestly unitary approach to virtual corrections 
an instructive one, especially in
application to the evolution of the color density
matrix for collective nuclear glue. 
For instance, when discussing the
unitarity properties of diffraction, one needs to split
the $q\bar{q}g$ contribution to diffractive cross section 
into the pQCD radiative correction to the low- to 
moderate-mass $q\bar{q}$ diffraction and excitation of 
the high-mass $q\bar{q}g$ states. 
Here the correct result for
the  high-mass $q\bar{q}g$ unitarity cut was found to be different
from what one would expect from the reinterpretation
of diffraction as deep inelastic scattering off pomerons
treated as a particle \cite{3POMglue}. Similar 
conclusions were reached by Braun who correctly noticed 
how applications of AGK rules 
depend on the form of the so-called triple-pomeron vertex.
(\cite{Braun3P,Braun2006}).

\begin{widetext}
\mbox{}
\begin{figure}[!h]
\begin{center}
\includegraphics[width = 4cm,angle=270]{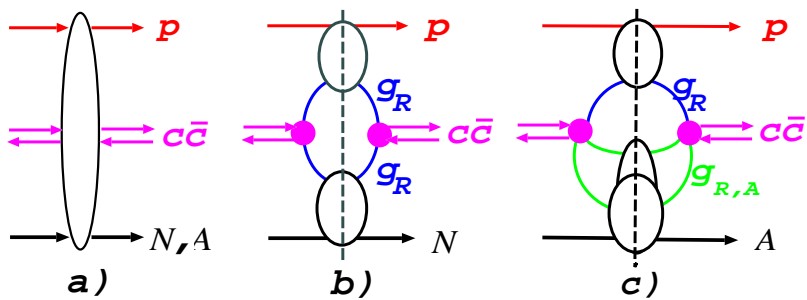}
\caption{\label{fig:KMcharm2}
(a) The Kancheli-Mueller diagram for the $c\bar{c}$ jet spectrum, 
(b) its explicit form in terms of the emission of  a $c\bar{c}$
pair off reggeized gluons $g_R$ in proton-nucleon collisions,
(d) the same as (b) for proton-nucleus collisions 
in terms of two cut nuclear pomerons $\Pom_{A,g}$
composed of effective nuclear reggeized gluons $g_{R,A}$
and the vertex of emission of a $c\bar{c}$ in a transition
from reggeized gluon $g_R$ to two effective nuclear
gluons $g_{R,A}$.}
\end{center}
\end{figure}
\mbox{}
\end{widetext}

Now, following our early discussion
in \cite{SingleJet}, we comment that the above multiregge form for 
mid-rapidity, Fig. \ref{fig:KMglue2}e, is quite likely 
to be an exclusive property of single gluon production,
and the multipomeron version,  Fig. \ref{fig:KMglue2}f,
is not an artefact.\footnote{Although a
direct connection between the nuclear
pomerons and  next-to-leading order BFKL pomeron is
not an obvious one, here 
we cite Fadin et al.
who in their recent study of the next-to-leading order BFKL 
stated that a rigorous proof of the multiregge 
factorization in elementary scattering 
can be carried through only for one-gluon
production \cite{FadinNew}.}
Indeed, a jet
can well be formed by a fixed-mass two-parton 
or higher multiplicity multiparton state.
As an illustration we look at the universality class of 
mid-rapidity open charm pair $c\bar{c}$ when the transverse
momentum much larger than the mass 
of the diparton. 
The factorization properties of mid-rapidity
$c\bar{c}$ production were studied in 
\cite{QuarkGluonDijet,Nonuniversality,Paradigm,SingleJet}.
The principal finding is a full fledged nonlinear
$k_\perp$-factorization for mid-rapidity $c\bar{c}$ 
production. Consequently, in the corresponding 
$3\to 3$ Kancheli-Mueller
diagram the exchange between the   $c\bar{c}$ jet and
the target nucleus can not be described by
the single nuclear pomeron. 

It is sufficient to 
consider the $g \to c\bar{c}$ excitation and Born
cross sections to the large $N_c$ approximation. 
Regarding the color properties, at large $N_c$ the $c\bar{c}$
pair is a composite gluon.
Let $z$ and $\bar{z}=1-z$
be a partition of the  $c\bar{c}$ jet momentum between
the quark and antiquark, $\bp$ the total transverse
momentum of the open charm jet, and $\bk$ the transverse
momentum of the quark with respect to the jet, so that
\beq
M_{\perp,c\bar{c}}^2 = {\bk^2+m_c^2 \over z(1-z)}+\bp^2. 
\label{eq:5.F.6}
\eeq
The differential form of Eq. (23) of Ref. \cite{SingleJet}
gives the free-nucleon spectrum
\bea 
&&{2(2\pi)^2 d\sigma_N(g^* \to c\bar{c}) 
\over dz d^2\bp d^2\bk}= 2\alpha_S P_{qg}(z)f(\bp)
\nonumber\\
&&\times 
[K(z\bp+\bk,\bk)+K((1-z)\bp-\bk,-\bk)],\nonumber\\
\label{eq:5.F.7}
\eea
Here a square of the vertex of emission of the $c\bar{c}$ pair from
a reggeized gluon, $ (c\bar{c})g_R g_R$ in Fig. \ref{fig:KMcharm2}b, 
is $\propto 
P_{qg}(z)[K(z\bp+\bk,\bk)+K((1-z)\bp-\bk,-\bk)]$.
The nuclear $c\bar{c}$ jet spectrum equals  
\bea 
&&{2(2\pi)^2 d\sigma_A(g^* \to Q\bar{Q}) 
\over d^2\bb dz d^2\bp d^2\bk}= \nonumber\\
&&=2\alpha_S P_{qg}(z)
\int d^2\bkappa_1   d^2\bkappa_2 \delta(\bp-\bkappa_1-\bkappa_2)\nonumber\\
&&\times
\Phi_g(\bb,x_A,\bkappa_1) \Phi_g(\bb,x_A,\bkappa_2)\nonumber\\
&&\times 
[K(z\bp+\bk-\bkappa_1,\bk)
\nonumber\\
&&+ K((1-z)\bp-\bk-\bkappa_1,-\bk)]
\label{eq:5.F.8}
\eea
and corresponds to the Kancheli-Mueller diagram of 
Fig. \ref{fig:KMcharm2}c with two cut nuclear pomerons 
$P_{A,g}$. Then Eq. (\ref{eq:5.F.8}) gives an explicit
form of the quark loop contribution to the fully 
cut triple-pomeron vertex, and a comparison with 
Eq. (\ref{eq:5.F.7}) suggests a simple correspondence between
this triple-pomeron vertex and the square of the $ (c\bar{c})g_R g_R$ 
vertex. Eq. (\ref{eq:5.F.8}) is the Born approximation,
the discussion of LL${1\over x}$ iterations 
in Sec. V.D suggests that upon the pQCD
radiative corrections $\Phi_g(\bb,x_A,\bkappa) \Phi_g(\bb,x_A,\bkappa_2)$
is substituted by the LL${1\over x}$ evolved
$\Phi_g(\bb,x,\bkappa) \Phi_g(\bb,x,\bkappa_2)$, although this
conjecture must be checked.

Finally, we briefly comment on a possible structure of the
related Kancheli-Mueller diagrams for digluon jets, a 
more detailed analysis will be reported elsewhere \cite{NSZ_AGK}. 
We recall that the mid-rapidity open charm production 
belongs to the universality class of nonlinear $k$-factorization 
which is free of coherent distortions of the dipole wave functions.
Also, at large $N_c$ it is a single-channel problem.
In contrast to that, intranuclear evolution of digluons 
is a full fledged multichannel non-Abelian problem, 
the pattern of nonlinear $k$-factorization depends
on the color state of digluons and there are coherent 
distortions of the color dipole wave function 
which change from one universality class to another
and depend on the depth in a nucleus at which digluons in color
representations higher than the adjoint representation
are excited \cite{Nonuniversality,Paradigm,GluonGluonDijet}.
The corresponding Kancheli-Mueller diagrams for digluons 
will be similar to Fig. \ref{fig:KMcharm2}c complemented
by uncut pomeron exchanges on both sides of the unitarity
cut, which describe the nuclear distortions of 
the corresponding partially cut multipomeron vertices. In the
universality class of digluons in higher color representations
these   partially cut multipomeron vertices must be
averaged over the depth in a nucleus.


\section{Summary and conclusions}

We reported a derivation of the transverse momentum dependent LPM
effect for leading jets in the beam fragmentation of proton-proton 
and proton-nucleus collisions. In conjunction with  
the LL${1\over x}$ evolution for the
collective nuclear unintegrated glue, our results offer a basis 
for a viable pQCD phenomenology of the LPM effect -- the
radiation driven quenching -- of leading
jets in $pp$ and $pA$ collisions in the finite energy range from
RHIC to LHC.

The first novelty of our work is
the derivation of the virtual radiative correction to the spectrum
of leading jets. The quenching of forward jets is caused by
radiation of hard gluons and, by virtue of the unitarity relation,
the opening of the radiation channel is followed by the
renormalization of the radiationless process. Our derivation
of this renormalization, i.e., of the virtual radiative correction,
is based on an explicit solution of the $s$-channel unitarity 
relation. While for the free-nucleon target both the real emission
and virtual radiative corrections are linear $k_\perp$-factorizable,
for nuclear targets the quantitative description of the LPM effect 
for forward dijets requires the full-fledged nonlinear
$k_\perp$-factorization. 

The second novelty of our work is an establishment of
the LPM effect-deconvoluted, i.e., the LL${1\over x}$
approximation, spectrum of leading partons as a
unique linear probe of the collective unintegrated nuclear glue
as defined through the excitation of coherent diffractive dijets
off nuclei. This must be contrasted to the manifestly nonlinear
$k_\perp$-factorization for all dijet observables and
the LPM effect for single-jet spectra. Although 
both real-emission and virtual radiative corrections to the leading
jet spectrum are described by nonlinear $k_\perp$-factorization,
our new finding is that this nonlinear $k_\perp$-factorization
for leading jets 
exactly matches the equally nonlinear $k_\perp$-factorization
for the  LL${1\over x}$ evolution of the nuclear glue.
The two nonlinear evolving observables prove to be 
proportional to each other: the leading jet production maps 
linearly the 
collective nuclear glue.
Our focus was on the impact of gluons radiated in the rapidity
space 
between the observed leading jet and the target nucleus, but all our
observations on the combined  effect of the VRC and REC are 
fully applicable to 
production of mid-rapidity jets too. 
Here our 
nonlinear $k_\perp$ factorization results suggest interesting reggeon
field theory properties of the Kancheli-Mueller diagrams
for multiparton jets. 

For the free-nucleon targets such a linear 
$k_\perp$-factorization for,  and the 
BFKL evolution property of, the leading jet spectrum 
are valid to all orders of  
the  LL${1\over x}$ BFKL evolution. While the above
specified definition of the collective nuclear  glue
in terms of coherent diffraction off nuclei
is viable to all orders of the LL${1\over x}$ evolution, 
our  proof of the linear $k_\perp$-factorization
for leading jets off nuclei is rigorous only to the first iteration
of the  LL${1\over x}$ evolution. Even this result is 
sufficient for the quantitative predictions of the quenching of
leading jets at RHIC because of the not so large energy of RHIC.
A similar description of the experimental data at the much higher 
energies of LHC requires several -- two to three --
steps of the LL${1\over x}$ evolution. We argue
that for such a small number of iterations, the treatment 
of heavy nuclei as a dilute uncorrelated gas of nucleons
is still viable, and in the finite energy range from RHIC 
to LHC, one can perform this evolution in the Balitsky-Kovchegov 
approximation. After such an evolution, our evaluation of the LPM
effect for leading jets will be fully applicable to a
comparison of $pp$ and $pA$ 
collisions at LHC. 

Our analysis of the nonlinear $k_\perp$-factorization
properties of the BK approximation revealed a remarkable feature:
the evolution equation can be cast in the form of fully
linear BFKL evolution for the collective nuclear glue complemented
by a nonlinear term which for hard gluons is a pure 
higher twist correction. To our opinion, this new form of 
the evolution is best suited for several iterations needed 
to bridge the energy range from RHIC to LHC.

We conclude by noting that one process which invites an application 
of the found linear
$k_\perp$-factorization for leading jets is the coherent
diffractive deep inelastic scattering scattering off nucleons 
and nuclei. The differential cross section of forward diffractive DIS
is a sum of differential cross section of elastic scattering
of multiparton Fock states of the photon \cite{NZ92,NZ94}.
As such, diffractive DIS is an inherently nonlinear process 
to be described by nonlinear $k_\perp$-factorization.
One can insist, though, on the linear color dipole representation 
of the diffractive cross section similar to that of 
inclusive DIS \cite{NZ94,NSZZdiffrValence}. Such a representation
would entail a certain nonlinear $k_\perp$-factorization form 
of the unintegrated glue in the pomeron. On the other hand,
motivated by the above discussion, one may try to define the
glue in the pomeron in terms of the transverse momentum
spectrum of leading (anti)quark jets in the high-mass diffractive
system.  The recent finding is that the two definitions of
the unintegrated glue in the pomeron would differ one from
another - one must be cautious with the unitarity cut
reinterpretation of the color dipole representation for diffraction
\cite{3POMglue}. 

\section*{Acknowledgments}

Thanks are due to A. Szczurek, B.G. Zakharov and V.R. Zoller for 
useful discussions.

\setcounter{equation}{0}
\renewcommand{\theequation}{A.\arabic{equation}}

\section*{Appendix: Still more linear form of the nonlinear evolution}

After we partly vindicated the utility  of the BK approximation for 
finite-energy evolution and heavy nuclei, we report here an
interesting alternative form of this equation
(the preliminary discussion was reported 
elsewhere \cite{NonlinearBFKL}), which offers
a fresh view at the distinction between the linear and
genuinely nonlinear components of the evolution equation
for the triplet-antitriplet dipoles
(\ref{eq:5.C.5}). An obvious drawback of this form
is a reference to the explicit Glauber-Gribov
form of the nuclear attenuation
factor. 
However, making use of the sum rule 
$\int d^2\bkappa \Phi(\bb,x_0,\bkappa)=1$, i.e.,
$\int d^2\bkappa \phi(\bb,x,\bkappa)=
1- \textsf{S}[\bb;\sigma_{q\bar{q}0}(x_0)]$,
one can  bring it into the form,
\bea {\partial \phi(\bb,x,\bp) \over \partial \log(1/x)} &&=
{\cal{K}}_{BFKL}\otimes \phi(\bb,x,\bp) 
\nonumber \\ && +
{\cal{Q}}[\phi](\bb,x,\bp) \, , 
\label{eq:Append.1}
\eea
where the nonlinear $k_\perp$-factorization
(quadratic)  functional ${\cal{Q}}[\phi]$ is
given by
\begin{widetext}
\bea {\cal{Q}}[\phi](\bb,x,\bp) &=& \int d^2\bq d^2\bkappa \phi(\bb,x,\bq)
\Big\{\Big[ K(\bp+\bkappa,\bp+\bq) - K(\bp,\bkappa+\bp) -
K(\bp,\bq+\bp)\Big] \nonumber\\
&\times&\phi(\bb,x,\bkappa) 
 - \phi(\bb,x,\bp) \Big[K(\bkappa,\bkappa+\bq+\bp) -
K(\bkappa,\bkappa+\bp)\Big] \Big\} \, .
\label{eq:Append.2}
\eea
\end{widetext}
The first iteration of (\ref{eq:Append.1}) with $x=x_0=x_A$ in the rhs
is exact. The so-defined  ${\cal{Q}}[\phi](\bb,x_0,\bp)$
is free of $\textsf{S}[\bb;\sigma_{0}(x_0)]$ and we
view it as a better candidate for finite-energy iterations
as discussed in Sec. V.E. 
If one would identify $\phi(\bb,x,\bp)$ with the exchange
by collective nuclear pomeron, then ${\cal{Q}}[\phi]$ describes
the fusion of two nuclear pomerons,  each of which is a nonlinear functional of,
and sums a multiple exchange by,
the free-nucleon pomerons.

In the case of the azimuthally symmetric $\phi(\bb,x,\bp)$, i.e., 
the  conformal spin zero in the $t$-channel, the
nonlinear $k_\perp$-factorization functional  ${\cal{Q}}[\phi]$ 
takes a remarkably simple form. In this case the azimuthal averaging
can be performed explicitly and leads to enormous simplifications. 
For instance, when  $\varepsilon^2$ can be neglected compared to
all the momenta, we make use of
\bea 
\Big \langle {\bkappa + \bp \over (\bkappa + \bp)^2} \Big
\rangle_{\bkappa} &=& {\bp \over \bp^2} \, \theta(\bp^2 -
\bkappa^2) \, ,
\label{eq:Append.3}
\eea
the related result for a finite $\varepsilon^2$ in the wave function
(\ref{eq:4.A.7}) is found in \cite{NZ92,NZsplit}, see also below. Here 
the subscript indicates the azimuthal angle of
the momentum the averaging is performed over. 

First we notice that for negligible small $\varepsilon^2=\mu_g^2$
\bea
&&K(\bp+\bkappa,\bp+\bq) - K(\bp,\bkappa+\bp) -
K(\bp,\bq+\bp)=\nonumber\\
&=& 2\Big[{\bp\over \bp^2}\cdot {\bkappa+\bp\over (\bkappa+\bp)^2}+
{\bp\over \bp^2}\cdot {\bq+\bp\over (\bq+\bp)^2}\nonumber\\
&-&
{\bkappa+\bp\over (\bkappa+\bp)^2}\cdot {\bq+\bp\over (\bq+\bp)^2}
- {1\over \bp^2}\Big].
\label{eq:Append.4}
\eea
Upon the averaging over the azimuthal angles of $\bkappa$ and $\bq$
we obtain
\begin{widetext}
\bea
&&\Big\langle \Big\langle K(\bp+\bkappa,\bp+\bq) -
K(\bp,\bkappa+\bp) - K(\bp,\bq+\bp) \Big\rangle
\Big\rangle_{\bq, \bkappa} = 
- {2 \over \bp^2} \theta(\bkappa^2 - \bp^2 ) \theta(\bq^2 -
\bp^2 )  \, .
\label{eq:Append.5}
\eea
\end{widetext}
In the second piece of (\ref{eq:Append.2}) we make first a judicious shift
of the integration variable $\bkappa$ in the convergent 
integral:
$K(\bkappa,\bkappa+\bq+\bp) - K(\bkappa,\bkappa+\bp)\Rightarrow 
K(\bkappa-\bp,\bkappa+\bq) - K(\bkappa,\bkappa+\bp)$. Next,
we perform the averaging over the azimuthal angles of $\bq$ and
the external variable $\bp$. There emerges the convergent,
and vanishing,
integral 
\beq
\int d^2\bkappa\Biggl[{1\over \bkappa^2}-{1\over (\bkappa+\bq)^2}\Biggr]=0.
\label{eq:Append.6}
\eeq
Upon this cancellation,  for negligible small $\varepsilon^2$,
\bea
&&\Big\langle \Big\langle  K(\bkappa-\bp,\bkappa+\bq) - K(\bkappa,\bkappa+\bp) 
\Big\rangle \Big\rangle_{\bp,\bq} =\nonumber\\
&=&
2\Big\langle \Big\langle {\bkappa\over \bkappa^2}\cdot 
{\bkappa+\bp\over (\bkappa+\bp)^2}-
{\bkappa-\bp\over (\bkappa-\bp)^2}\cdot {\bkappa+\bq\over (\bkappa+\bq)^2}
\Big\rangle \Big\rangle_{\bp,\bq}\nonumber\\
&=& {2 \over \bkappa^2} \theta(\bkappa^2-\bp^2) \theta(\bq^2-\bkappa^2) \,
\label{eq:Append.7}
\eea
Putting it all together the quadratic functional ${\cal{Q}}[\phi]$
assumes the very simple form,
\begin{widetext}
\bea {\cal{Q}}[\phi](\bb,x,\bp) &=& - {2 {\cal K}_0 \over \bp^2}
\int d^2\bq d^2\bkappa \phi(\bb,x,\bq)\phi(\bb,x,\bp) \theta(\bq^2-\bp^2)
\theta(\bkappa^2-\bp^2) \nonumber\\
&&- 2 {\cal K}_0 \phi(\bb,x,\bp) \int{d^2\bkappa \over \bkappa^2} \int d^2\bq
\phi(\bb,x,\bq) \theta(\bkappa^2-\bp^2) \theta(\bq^2-\bkappa^2) 
\\
&&= -{2  {\cal K}_0 \over \bp^2} \Biggl[ \int_{\bp^2}^\infty d^2 \bq
\phi(\bb,x,\bq) \Biggr]^2 - 2 {\cal K}_0 \phi(\bb,x,\bp) \int_{\bp^2} ^\infty {d^2\bkappa
\over \bkappa^2} \int_{\bkappa^2} ^\infty d^2\bq \phi(\bb,x,\bq),\nonumber
\label{eq:Append.8}
\eea
\end{widetext}
valid for the transverse momenta larger than the infrared regularization
$\mu_g$, for the effects of finite $\mu_g$ see equations (\ref{eq:Append.15} 
- \ref{eq:Append.19}) 
below.

It is instructive to 
reformulate it in terms of the collective nuclear glue per
bound nucleon, ${\cal F}_A(\bb,x,\bp)$:
\bea
\phi(\bb,x,\bp) ={2\pi \alpha_S(\bp^2) \over N_c p^4}T(\bb){\cal F}_A(\bb,x,\bp).
\label{eq:Append.9}
\eea
One readily obtains
\bea
&&{\partial {\cal F}_A(\bb,x,\bp) \over \partial\log{1\over x}}
={\partial^2  G_A(\bb,x,\bp) \over 
\partial\log \bp^2 \partial\log{1\over x}}
\nonumber\\
&=& {\cal K}_{BFKL} \otimes {\cal F}_A(\bb,x,\bp) 
\nonumber \\
&-&
{4\pi {\cal K}_0 \over  N_c } \alpha_S(\bp^2) T(\bb)
\Biggl\{\bp^2\Bigl[ \int_{\bp^2}^\infty {d^2 \bq\over q^4}
 {\cal F}_A(\bb,x,\bq) \Bigr]^2 \nonumber\\
&+&
  {\cal F}_A(\bb,x,\bp)\int_{\bp^2} ^\infty {d^2\bkappa
\over \bkappa^2} \int_{\bkappa^2} ^\infty{ d^2\bq \over q^4}
{\cal F}_A(\bb,x,\bq)\Biggr\},\nonumber\\
\label{eq:Append.10}
\eea
The fusion term is of manifestly negative, i.e., shadowing, sign.
As such, Eq. (\ref{eq:Append.2}) is the better candidate for the definition
of of the triple-pomeron vertex for nuclear pomerons than Eq. (\ref{eq:5.C.5}).
Such a representation for the 
fusion term  is a new result, it is exact for all  momenta, 
both above the saturation scale,
 $\bp^2\gsim Q_A^2$, 
and under the saturation
scale, $\bp^2\lsim Q_A^2$, provided that $\bp^2 \gsim \mu_g^2$. 

Now we shall argue that at large momenta above the
the saturation scale the fusion term is of pure higher 
twist. We focus on $x=x_0$,
where one can rely upon the 
saturation, antishadowing and hard momentum properties of
$\phi(\bb,x_0,\bp)$ found in \cite{NSSdijet,Nonlinear,SingleJet}. 
Here we need the large-momentum approximation
\bea
\phi(\bb,x_0,\bp) &\approx&{1\over 2} T(\bb) f(x_0,\bp) \propto{ 1 \over (\bp^2)^\gamma}\nonumber\\
\label{eq:Append.11}
\eea
where the exponent $\gamma\approx 2$, see Eq. (\ref{eq:2.B.6}). 
Making the explicit use of
this asymptotic behavior, we 
obtain 
\bea
&&{1 \over \bp^2}\Biggl[ \int_{\bp^2}^\infty d^2 \bq
\phi(\bb,x_0,\bq) \Biggr]^2 \approx \nonumber\\
&&\approx
\phi(\bb,x_0,\bp)\int_{\bp^2}^\infty {d^2\bkappa
\over \bkappa^2} \int_{\bkappa^2}^\infty d^2\bq \phi(\bb,x_0,\bq)\nonumber\\
&&
\approx {\pi^2\over (\gamma -1)^2}
\bp^2 \phi^2(\bb,x_0,\bp^2).
\label{eq:Append.12}
\eea
The fusion term takes the form of the boundary condition glue times 
the higher twist factor,
\bea &&{\cal{Q}}
[\phi](\bb,x_0,\bp) \approx \nonumber\\
&-&{4\pi^2{\cal K}_0\over (\gamma -1)^2}\phi(\bb,x_0,\bp)\cdot
\{\bp^2 \phi(\bb,x_0,\bp)\}\nonumber\\
&\approx & {2\pi \alpha_S(\bp^2) \over (\gamma -1)^2
\alpha_S(Q_A^2(\bb,x_0))G(x_0,Q_A^2(\bb,x_0))}
\nonumber\\
&\times&
{Q_A^2(\bb,x_0)\over \bp^2} \cdot
\phi(\bb,x_0,\bp^2), 
\label{eq:Append.13}
\eea
where the scale for the higher twist is set by the saturation
scale. In terms of the collective glue per bound nucleon,
this amounts to
\bea
&&{\partial {\cal F}_A(\bb,x,\bp) \over \partial\log{1\over x}}
={\cal K}_{BFKL} \otimes {\cal F}_A(\bb,x_0,\bp) 
\nonumber\\
&-&
{8{\cal K}_0 \pi^3 \alpha_S(\bp^2) T(\bb)
\over (\gamma-1)^2 N_c \bp^2}  {\cal F}_A^2(\bb,x_0,\bp).
\nonumber \\
\label{eq:Append.14}
\eea
In contrast to the exact integral form (\ref{eq:Append.10}) 
of the fusion term,
the local in the gluon momentum form (\ref{eq:Append.14}) 
is an approximation valid for the specific
parameterization (\ref{eq:Append.11}), for alternative 
forms of the fusion correction see  \cite{GLR,MuellerQiu}.
The further discussion of (\ref{eq:Append.10})
and (\ref{eq:Append.14}) goes beyond the scope
of this communication. We restrict ourselves to the comment that
the fusion term does 
not exhaust the nuclear higher
twist corrections --  we recall that also the boundary condition 
${\cal F}_A(\bb,x_0,\bp^2)$ contains a substantial
and very similar, but positive valued,
i.e., antishadowing, linear BFKL-evolving nuclear higher twist 
correction proportional to ${\cal F}_A(\bb,x_0,\bp^2)
G_A(\bb,x_0,Q^2)T(\bb)/\bp^2$
\cite{NSSdijet,NSSdijetJETPLett,Nonlinear}. The shadowing and
antishadowing nuclear higher twist corrections both have their origin
in unitarity, the impact of the antishadowing correction was not
given a proper discussion earlier. 

For the sake of completeness, we show the integral
form of the fusion term in Eq.~(\ref{eq:Append.10}) subject to
the infrared regularization (\ref{eq:5.A.3}), i.e. 
in all wave functions, $K(\bp,\bq)$ and
${\cal K}_{BFKL}$ we put
$\varepsilon^2=\mu_g^2$.  Following
\cite{NZ92,NZsplit} we can write 
\bea 
\Big \langle {\bkappa + \bp \over (\bkappa + \bp)^2+\mu_g^2} \Big
\rangle_{\bkappa} &=& {\bp \over \bp^2+\mu_g^2} \, \theta(\bp^2,\bkappa^2,\mu_g^2) ,\nonumber\\
\label{eq:Append.15}
\eea
where
\bea
 &&\theta(\bp^2,\bkappa^2,\mu_g^2)  = {\bp^2+\mu_g^2\over 2\bp^2}
\nonumber\\
&&\times
\left(1+ {\bp^2-\mu_g^2-\bkappa^2 \over 
\sqrt{ ( \bp^2-\mu_g^2-\bkappa^2)^2+4\bp^2\mu_g^2}}\right).\nonumber\\
\label{eq:Append.16}
\eea
Eqs.~(\ref{eq:Append.5}) and ~(\ref{eq:Append.5}) take the form 
\begin{widetext}

\bea
&&\Big\langle \Big\langle K(\bp+\bkappa,\bp+\bq) -
K(\bp,\bkappa+\bp) - K(\bp,\bq+\bp) \Big\rangle
\Big\rangle_{\bq, \bkappa} = 
\nonumber \\ &=&
- {2 \bp^2\over (\bp^2+\mu_g^2)^2} [1- \theta(\bp^2,\bkappa^2,\mu_g^2)]
 [1- \theta(\bp^2,\bq^2,\mu_g^2)],
\label{eq:Append.17}
\eea
\bea
\Big\langle \Big\langle  K(\bkappa-\bp,\bkappa+\bq) - K(\bkappa,\bkappa+\bp) 
\Big\rangle \Big\rangle_{\bp,\bq} = {2 \bkappa^2 \over (\bkappa^2 +\mu_g^2)^2}
\theta(\bkappa^2,\bp^2,\mu_g^2)
 [1- \theta(\bkappa^2,\bq^2,\mu_g^2)].
\label{eq:Append.18}
\eea
The infrared-regularization modifies the evolution equation (\ref{eq:Append.10})
as follows:
\bea
{\partial {\cal F}_A(\bb,x,\bp) \over \partial\log{1\over x}}
&=& {\cal K}_{BFKL} \otimes {\cal F}_A(\bb,x,\bp) 
\nonumber \\
&-&
{4\pi {\cal K}_0 \over  N_c } \alpha_S(\bp^2) T(\bb)
\Biggl\{{p^6\over (\bp^2+\mu_g^2)^2}\Bigl[ \int {d^2 \bq\over q^4}
  [1- \theta(\bp^2,\bq^2,\mu_g^2)] {\cal F}_A(\bb,x,\bq) \Bigr]^2 \nonumber\\
&+&
  {\cal F}_A(\bb,x,\bp)\int \int{d^2\bkappa d^2\bq 
\over (\bkappa^2+\mu_g^2)^2 q^4} \theta(\bkappa^2,\bp^2,\mu_g^2)
 [1- \theta(\bkappa^2,\bq^2,\mu_g^2)]
{\cal F}_A(\bb,x,\bq)\Biggr\}.\nonumber\\
\label{eq:Append.19}
\eea
\end{widetext}

\end{document}